\documentclass[reprint,pre,amsmath,amssymb,aps,floatfix,superscriptaddress]{revtex4-2}
\usepackage{graphicx,bm,dsfont}
\usepackage{amsmath,amsfonts}
\usepackage{color,xcolor,comment}
\usepackage{amssymb} 

\usepackage{tikz}

\usepackage{hyperref}

\begin{document}

\title{Imprinting Macroscopic Fracture during Gelation:\\ A Mechanism for Tuning Colloidal Gels}

\author{Wilbert J. Smit}
\affiliation{ENSL, CNRS, Laboratoire de physique, F-69342 Lyon, France}
\author{Thomas Gibaud}
\affiliation{ENSL, CNRS, Laboratoire de physique, F-69342 Lyon, France}
\affiliation{Department of Polymer Engineering, IPC, University of Minho, Guimarães, 4804-533 Portugal}
\author{Sébastien Manneville}
\affiliation{ENSL, CNRS, Laboratoire de physique, F-69342 Lyon, France}
\affiliation{Institut Universitaire de France (IUF)}
\author{Thibaut Divoux}
\affiliation{ENSL, CNRS, Laboratoire de physique, F-69342 Lyon, France}

\begin{abstract}
Colloidal gels form through the sol-gel transition of attractive particle suspensions, where local aggregation leads to a space-spanning network with solid-like properties. Their microstructure and mechanical properties are highly sensitive to external perturbations, which can substantially alter the pathway of network formation. Here, we investigate how nonlinear oscillatory shear affects the sol-gel transition of colloidal silica suspensions. Using large-amplitude oscillatory shear (LAOS), we vary both the strain amplitude and the duration of oscillatory forcing, varying between one and two times the gelation time. We find that sufficiently large strain amplitudes, or prolonged exposure to oscillations in the nonlinear regime, alter irreversibly the gel properties: the storage modulus $G'$ decreases while its frequency dependence remains unchanged. In contrast, the loss modulus $G''$, which decreases monotonically with frequency under quiescent gelation, exhibits an upturn at high frequencies when the gel is formed under strong oscillatory shear. The viscoelastic spectra of gels formed under quiescent conditions are well captured by a fractional Maxwell model, while gels formed under LAOS require an additional fractional element to account for damage-induced dissipation. Rheo-imaging experiments corroborate this interpretation by revealing the growth of cracks in gels formed under LAOS. We further show that these gels display a progressively more ductile nonlinear response for prolonged exposure to LAOS during gelation. These results demonstrate that the interplay between nonlinear shear and gelation can permanently imprint a macroscopic fracture pattern into colloidal gels, offering a route to tune their viscoelastic properties.
\end{abstract}

\maketitle

\section{Introduction}

Colloidal gels are soft solids formed when attractive particles in suspension aggregate into a percolated network, which endows the material with solid‐like mechanical response, even at a low particle volume fraction \cite{Bonn:2017,Royall:2021}. The sol-gel transition has been extensively studied both experimentally and theoretically under quiescent conditions, i.e., in the absence of any imposed flow or any external perturbation \cite{rueb:1998,Bergenholtz:2003,Varga:2015,Rouwhorst:2020,MorletDecarnin:2023}. Particles in suspensions that display attractive interactions tend to aggregate to form clusters, or denser regions, when following a phase-separation scenario, which in turn serve as building blocks for larger-scale structures, eventually forming a space-spanning network. The gel point, which corresponds to the moment when a percolated network appears, is characterized by a peculiar power-law mechanical response, intimately connected to the hierarchical structure of the gel backbone \cite{Aime:2018,Keshavarz:2021}. 
Beyond the gel point, the network tends to show heterogeneous fractal structures whose mesh size, connectivity, and strand thickness depend on particle concentration and interaction strength \cite{Johnson:2019}. In that context, the establishment of a quantitative link between the gel microstructure and its macroscale mechanical response remains an active research area \cite{Whitaker:2019,Kao:2022,Bantawa:2023,Richard:2025}.

In practice, however, sol-gel transition rarely occurs under quiescent conditions: external perturbations that arise during processing (e.g. casting, molding, or printing) can strongly affect the gel microstructure \cite{Fournier:2024}. These effects have been explored in a broad variety of experimental and numerical studies involving external perturbations such as continuous and oscillatory shear \cite{Hanley:1999,Mokhtari:2008,Koumakis:2015,Moghimi:2017,Sudreau:2022,Muzny:2023,Seljelid:2024,Bhaumik:2025a,Bhaumik:2025b}, ultrasonic waves \cite{Gibaud:2020,Dages:2021,SaintMichel:2022}, and electromagnetic fields \cite{Gadige:2018,Semwal:2022,Munteanu:2025}. Among these stimuli, external shear has been the most studied in the last 20 years. 

Flow at a constant shear rate in a suspension of attractive particles affects the aggregate size distribution, re-structuring smaller aggregates and fragmenting larger ones \cite{Selomulya:2001,Selomulya:2002,Eggersdorfer:2010,Lieu:2016,Bauland2024}. More generally, the impact of flow on a colloidal gel depends on the balance between hydrodynamic forces and interparticle adhesion. When shear stress exceeds the adhesive forces, the gel network is disrupted, whereas weaker flows primarily restructure the existing network. This balance is often quantified through the Mason number (Mn), a dimensionless number that compares viscous stresses to interparticle adhesive stresses \cite{Varga:2018,Varga:2019,Das:2021,Bauland2025}. For $\rm Mn<1$, external shear can reorganize the gel microstructure without complete disruption, typically leading to anisotropic textured networks. In contrast, for $\rm Mn>1$, the network is progressively eroded and eventually fully rejuvenated into a suspension of well-dispersed particles \cite{Jamali:2020}. 

In this framework, recent work has emphasized how the flow history of a colloidal gel --particularly shear applied beyond the gel point-- can leave persistent structural imprints that strongly influence its subsequent rheological properties \cite{Schwen:2020,Das:2022,Sudreau:2023}. Such ``structural memory" involves non-equilibrium rearrangements that alter network topology, cluster morphology, and connectivity in ways that are not recoverable (without applying a large shear with $\rm Mn \gg 1$). However, the outcome of these memory effects is not universal. In depletion gels, large shear rates homogenize the microstructure and reinforce elasticity \cite{Koumakis:2015}, while in carbon black and boehmite gels, strengthening occurs instead at low shear, linked to cluster interpenetration or shear-induced anisotropy \cite{Dages:2022,Sudreau:2022,Sudreau:2022b,Sudreau:2023}. In contrast, rod-based gels typically weaken under comparable conditions \cite{Das:2022}. This diversity of responses underscores the subtle interplay between particle interactions, microstructural anisotropy, and gel mechanics.
 
To disentangle these competing effects, we investigate gels formed by adhesive silica nanoparticles, a minimal model system where particle interactions are isotropic and tunable. This system enables a systematic test of how oscillatory forcing during gelation controls microstructure and mechanics. Our protocol consists in applying an oscillatory strain of amplitude $\gamma_{\rm OS}$ for a duration $\mathcal{T}_{\rm OS}$ throughout gelation, before switching to the linear deformation regime to monitor the subsequent evolution of the gel viscoelastic properties in a non-destructive way. This approach enables us to systematically investigate how the oscillatory shear applied during gelation influences network formation and the long-term mechanical properties of the resulting gel. 

We identify a critical duration $\mathcal{T}_c \simeq 1.45\,t_g$ (for $\gamma_{\rm OS}=40\%$) and a critical strain $\gamma_c \simeq 20\%$ (for $\mathcal{T}_{\rm OS}=2\,t_g$) beyond which the gels irreversibly damage due to the nucleation of macroscopic cracks, as confirmed by direct imaging in a transparent Couette cell. Remarkably, despite their apparent complexity, these fracture patterns leave a simple rheological fingerprint on the gel linear viscoelastic properties: $G'$ decreases while keeping the same frequency dependence, whereas $G''$ acquires a high-frequency upturn. Although quiescent gels are well described by the standard Fractional Maxwell model, an additional fractional element is needed to construct a Generalized Fractional Kelvin--Voigt model that captures damage-induced dissipation. Our study thus provides a simple modeling of the linear viscoelastic properties of silica gels whose microstructure is tuned by large-amplitude oscillatory shear that leads to crack formation.
By further examining the nonlinear response of gels formed under different oscillation amplitudes, we show that large amplitude oscillations applied during gelation not only weaken the elastic modulus but also alter the failure mode: the gels are all the more ductile when the amplitude of the imposed oscillatory strain is large. This ductility most likely originates from the pre-existing fracture pattern imprinted during gelation, which relaxes stress localization and delays catastrophic yielding.

The manuscript is organized as follows: Section~\ref{sec:rheology} introduces the materials and methods. Section~\ref{sec:duration} introduces the impact of the oscillatory time $\mathcal{T}_{\rm OS}$ on the long-term viscoelastic spectrum, while Section~\ref{sec:amplitude} reports on the impact of $\gamma_{\rm OS}$. Section~\ref{sec:rheoOptics} reports on direct optical observations performed during gelation under large amplitude oscillatory shear, revealing the formation of a crack pattern, which is further examined in light of the intracycle analysis of the shear stress and strain waveforms in Sect.~\ref{sec:Intracycle}. Finally, we probe the nonlinear response of gels formed under various oscillation amplitudes in Sect.~\ref{sec:Strainsweep}, unraveling that gels formed under large amplitude oscillation display a more ductile nonlinear response.

\begin{figure}[t]
\centering
\includegraphics[width=8.6 cm]{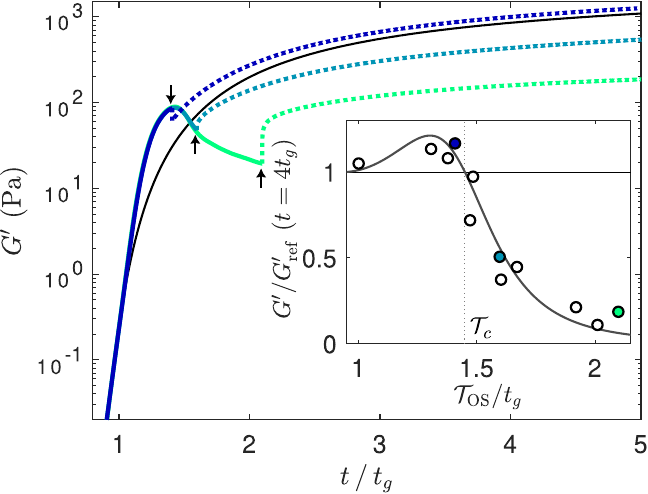} 
 \caption{Elastic modulus $G'$ as a function of normalized time $t/t_g$ during gelation, where $t_g=3590\pm190~\rm s$ is the gelation time determined by SAOS at $\omega=2\pi~\rm rad\ s^{-1}$. Each experiment was performed on a fresh sample, first under LAOS with amplitude $\gamma_{\rm OS}=40\%$ for a duration $\mathcal{T_{\rm OS}}$, and subsequently under SAOS with amplitude $\gamma_0=0.5\%$ for the rest of the experiment. Such a gelation under oscillatory shear was repeated for different $\mathcal{T_{\rm OS}}$ values: $\mathcal{T_{\rm OS}}/t_g=1.4$ (blue), $1.6$ (cyan), and $2.1$ (green). Vertical arrows mark the time at which the strain amplitude is switched from $\gamma_{\rm OS}=40\%$ to $\gamma_0=0.5\%$. The black curve corresponds to $G'_{\rm ref}(t)$, the gelation conducted entirely under SAOS with amplitude $\gamma_0=0.5\%$. Inset: normalized modulus $G'(t=4t_g)/G'_{\rm ref}(t=4t_g)$ as a function of $\mathcal{T_{\rm OS}}/t_g$. The vertical dotted line marks the critical duration $\mathcal{T}_c \simeq 1.45\,t_g$ beyond which large-amplitude oscillations irreversibly affect the gel linear viscoelastic properties. The solid black curve is a guide to the eye.}
\label{fig:evolutionLAOS}
\end{figure}

\begin{figure*}[t]
\centering
\includegraphics[width=\textwidth]{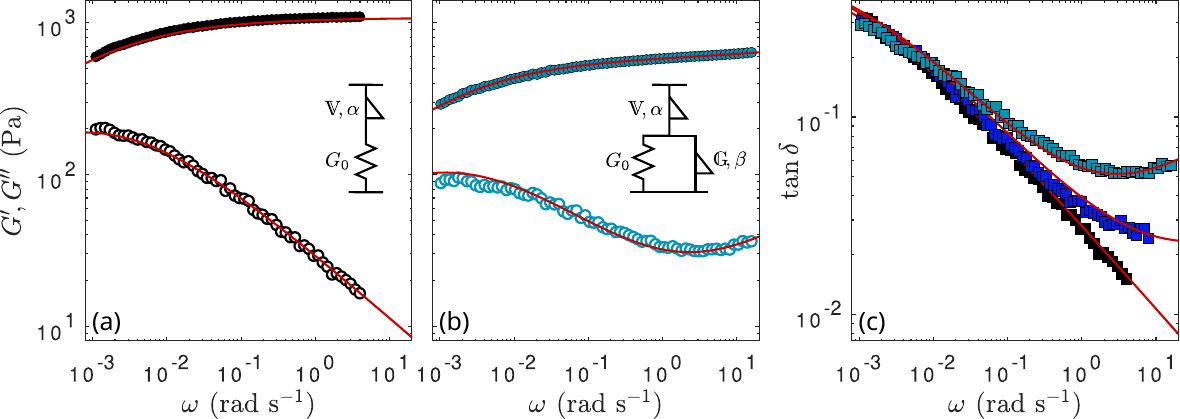} 
\caption{Linear viscoelastic spectra measured at long time ($t=5\,t_g$) for silica gels formed (a) under SAOS ($\gamma_0=0.5\%$), and (b) under LAOS with amplitude $\gamma_{\rm OS}=40\%$ applied during $\mathcal{T_{\rm OS}}=1.6\,t_g$. (c) Loss tangent $\tan \delta =G''/G'$ for the spectra shown in (a) and (b), together with an additional gel formed under LAOS at $\gamma_{\rm OS}=40\%$ for $\mathcal{T_{\rm OS}}=1.5\,t_g$. Colors in (c) are consistent with those used in (a) and (b). The viscoelastic spectrum of the silica gel formed under SAOS is fitted by a Fractional Maxwell model [see red curves in (a) and (c), and sketch in (a)]. The viscoelastic spectra of the gels formed under LAOS are fitted by a Generalized Fractional Kelvin--Voigt model [see red curves in (b) and (c), and sketch in (b)]. Fit parameters are reported as a function of $\mathcal{T_{\rm OS}}$ in Fig.~\ref{fig:40pctFitParams}.}
\label{fig:fig2}
\end{figure*}

\section{Materials and methods}

\subsection{Preparation of silica gels}
\label{sec:SilicaGel}
The samples consisted of suspensions of amorphous silica particles (Ludox\textregistered~HS-40, Sigma--Aldrich). The particles had a nominal diameter of $12~\rm nm$, as reported by the manufacturer and confirmed by dynamic light scattering on dilute samples (mean hydrodynamic diameter of $13.0\pm0.5$~nm). The 40~wt.\% stock solution was used as received without further purification, and diluted in aqueous NaCl to a final ionic strength of 0.61~M and a silica volume fraction of $0.05$. The added salt screens electrostatic repulsion between negatively charged silica particles, thereby promoting aggregation. Such salt-induced gelation of Ludox suspensions has been extensively studied, both in terms of microstructure and rheology, making this a well-established model system for colloidal gelation \cite{Trompette:2003,Tourbin:2008,Cao:2010,Vanderlinden:2015,Sogaard:2018,Keshavarz:2021}. \textcolor{black}{To avoid irreversible aging or particle sintering known to occur in silica suspensions under screened electrostatic conditions \cite{Bonacci:2020,Bonacci:2022}, each gelation experiment was performed on a freshly prepared sample.}
In the present study, the gelation time \textcolor{black}{$t_g$ is operationally defined as the crossover of the elastic and viscous moduli measured at a fixed frequency, and it} was observed to be $t_g=3590\pm190$~s irrespective of the amplitude $\gamma_{\rm OS}$ of the oscillatory forcing (see Appendix~\ref{sec:GelationTime} for the time evolution of both $G'$ and $G''$ during gelation and the determination of $t_g$).

\subsection{Rheometry and in-situ imaging}
\label{sec:rheology}
In this study, two different rheometers were used. A stress-controlled AR-G2 rheometer (TA Instrument) was used with a transparent Couette cell made of smooth polymethylmethacrylate, with inner radius $r_i=24~\rm mm$, outer radius $r_o=25~\rm mm$, and rotor height 58~mm. 
Optical images of the sample were acquired through the transparent outer wall of the Couette cell using a Pentax K-70 camera, with a pixel resolution of 3.2~$\mu$m at an acquisition rate of $24~\rm Hz$. A strain-controlled ARES-G2 (TA) was equipped with a cylindrical double gap stainless steel geometry, with an inner cup radius of $13.9~\rm mm$, an inner bob radius of $14.7~\rm mm$, an outer bob radius $16~\rm mm$, an outer cup radius of $17~\rm mm$, and an immersion height of $51~\rm mm$.
All experiments were carried out at 20$^\circ$C. The results presented in Sections~\ref{sec:duration}--\ref{sec:rheoOptics} were obtained with the stress-controlled setup, while those in Sections~\ref{sec:Intracycle}--\ref{sec:Strainsweep} were obtained with the strain-controlled rheometer.
In all cases, we used a gap distance of 0.7~mm between the bottom of the rotor and the base of the cup.

\section{Results}

\subsection{Impact of large-amplitude oscillation duration on gel elastic properties}
\label{sec:duration}

We first report on the impact of the duration of large amplitude oscillatory shear (LAOS) on the sol-gel transition. In practice, oscillations of amplitude $\gamma_{\rm OS}=40\%$ are imposed at a frequency $\omega = 2\pi~\rm rad\ s^{-1}$ over a duration $\mathcal{T_{\rm OS}}$ that was varied between $t_g$ and $2.1\, t_g$, where $t_g$ denotes the gelation time measured under quiescent conditions (see Sect.~\ref{sec:SilicaGel} and Appendix~\ref{sec:GelationTime}). For $t> \mathcal{T}_{\rm OS}$, the strain amplitude is switched to a smaller amplitude $\gamma_0=0.5\%$, within the linear deformation regime, to monitor the evolution of the gel viscoelastic properties in a non-perturbative fashion.

Figure~\ref{fig:evolutionLAOS} shows the evolution of the storage modulus $G'$ during the salt-induced gelation of the colloidal silica suspension measured by small-amplitude oscillatory shear (SAOS) tests under quiescent conditions (see black curve for which $\gamma_{\rm OS}=\gamma_0=0.5\%$). In comparison, salt-induced gelations under LAOS of strain amplitude $\gamma_{\rm OS} = 40\%$ applied for various durations $\mathcal{T}_{\rm OS}$ are colored. 
When large-amplitude oscillations are applied, the apparent early-time growth of $G'$ is faster than under quiescent conditions. This acceleration is consistent with recent in situ rheo-SAXS studies under continuous shear, which showed that the stress build-up intensifies with increasing shear rate due to shear-enhanced aggregation kinetics \cite{Seljelid:2024}. However, the early stiffening observed in $G'(t)$ is only transient: beyond a critical oscillation duration $\mathcal{T}_c \simeq 1.45\,t_g$, the subsequent growth of $G'$ is hindered: $G'(t)$ shows an overshoot, while the gel remains significantly weaker than its quiescent counterpart. 
This trend is summarized in the inset in Fig.~\ref{fig:evolutionLAOS}, where the long-time elastic modulus measured at an arbitrary time $t = 4\,t_g$ and normalized by the quiescent value is plotted as a function of the oscillation duration, expressed in units of the gelation time $t_g$. The data reveal a clear crossover at $\mathcal{T}_c$: below this threshold, the long-time elastic modulus is not significantly altered by LAOS, whereas exposure to oscillatory shear for a duration $\mathcal{T_{\rm OS}}>\mathcal{T}_c$ leads to a pronounced reduction in elasticity. These first results demonstrate that large-amplitude oscillations, depending on their duration relative to the intrinsic gelation time, can either promote or impair network formation, and thereby potentially tune the final mechanical properties of the gel.

\begin{figure} 
\centering
\includegraphics[width=\linewidth]{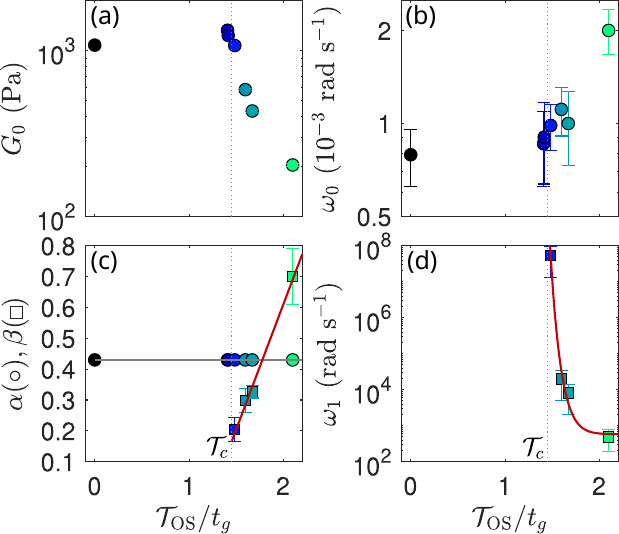}
\caption{Model parameters extracted from fits to fractional rheological models of the long-time viscoelastic spectra ($t=5\,t_g$, see Fig.~\ref{fig:fig2}) as a function of the oscillation duration $\mathcal{T_{\rm OS}}$ for a fixed LAOS amplitude $\gamma_{\rm OS}=40\%$ : (a) $G_0$, (b) characteristic frequency $\omega_0=(G_0/\mathbb{V})^{1/\alpha}$, (c) exponents $\alpha$ and $\beta$, and (d)~characteristic frequency $\omega_1=(G_0/\mathbb{G})^{1/\beta}$. Spectra measured on gels exposed to LAOS for $\mathcal{T_{\rm OS}}<\mathcal{T}_\mathrm{c}$ were fitted with the Fractional Maxwell model [Eq.~\eqref{eq:FMM}, inset in Fig.~\ref{fig:fig2}(a)], while those exposed to LAOS for $\mathcal{T}_{\rm OS}>\mathcal{T}_\mathrm{c}$ were fitted with the Generalized Fractional Kelvin--Voigt model for [Eq.~\ref{eq:GFKVM}, inset in Fig.~\ref{fig:fig2}(b)]. The vertical dotted line indicates the critical duration of oscillation $\mathcal{T}_c$ beyond which LAOS with amplitude $\gamma_{\rm OS}=40\%$ significantly alters the long-term linear viscoelastic properties of the gel. Error bars denote 95\% confidence intervals from the fit. In (c), the horizontal gray line marks the fixed value $\alpha=0.43$, imposed to all fits, irrespective of $\mathcal{T}_{\rm OS}$. In (c) and (d), red curves serve guide for the eye.}
\label{fig:40pctFitParams}
\end{figure}

\begin{figure*} 
\centering
\includegraphics[width=\textwidth]{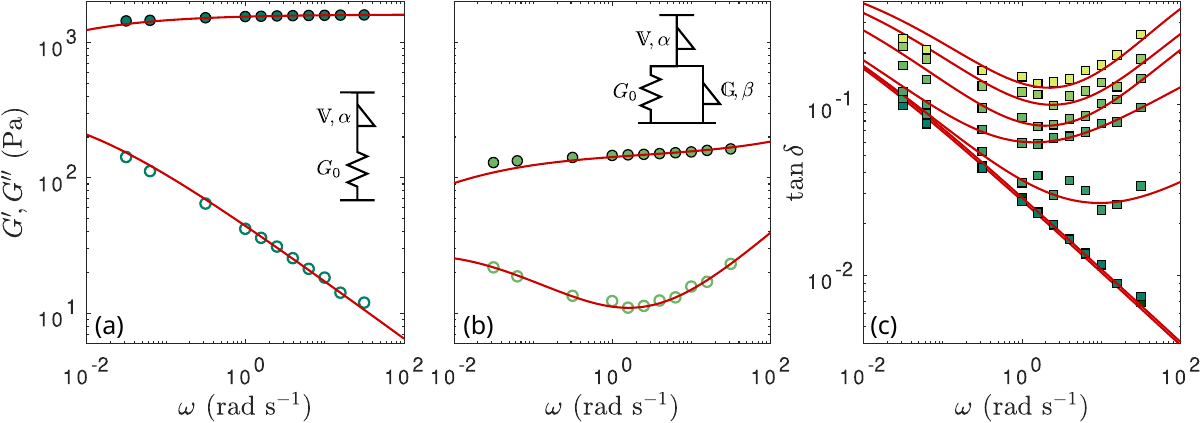}
\caption{Linear viscoelastic spectra measured at long time ($t=5\,t_g$) for silica gels formed (a) under SAOS with amplitude $\gamma_{\rm OS}=10\%$ and (b) under LAOS with amplitude $\gamma_{\rm OS}=40\%$ over the same duration $\mathcal{T_{\rm OS}}=2\,t_g$. (c) Loss tangent $\tan \delta =G''/G'$ vs. $\omega$ measured at $t=5\,t_g$, after the sample was exposed to oscillations of strain amplitude $\gamma_{\rm OS}=1, 10, 25, 30, 40, 50,$ and $80\%$ (from dark to bright) for $\mathcal{T_{\rm OS}}=2\,t_g$. The red curves correspond to the best fits of the data using a Fractional Maxwell model for $\gamma_{\rm OS} <\gamma_{c} \simeq 20\%$ [see sketch in (a)], and using a Generalized Fractional Kelvin--Voigt model for $\gamma_{\rm OS} > \gamma_{c} $ [see sketch in (b)]. Fit parameters are reported as a function of $\gamma_{\rm OS}$ in Fig.~\ref{fig:fractionalparameters}.}
\label{fig:spectra_amplitude}
\end{figure*}

\begin{figure} 
\centering
\includegraphics[width=\linewidth]{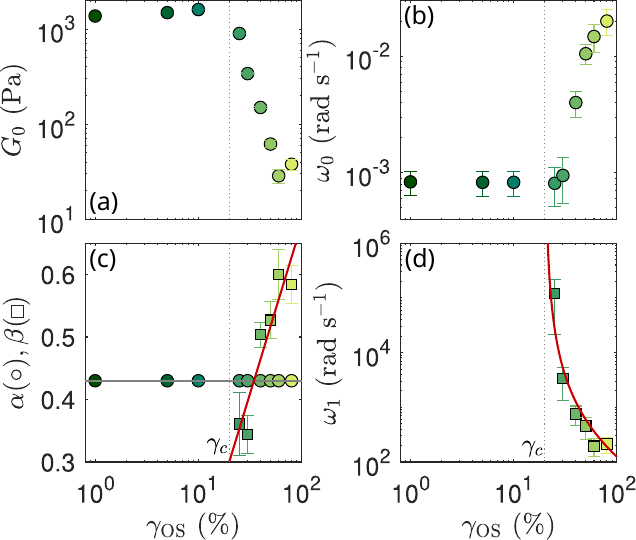}
\caption{Model parameters extracted from fits to fractional models of the long-time viscoelastic spectra ($t=5\,t_g$, see Fig.~\ref{fig:spectra_amplitude}) as a function of the strain amplitude $\gamma_{\rm OS}$ for a fixed oscillation duration $\mathcal{T}_{\rm OS}=2\,t_g$: (a) $G_0$, (b) characteristic frequency $\omega_0=(G_0/\mathbb{V})^{1/\alpha}$, (c) dimensionless exponents $\alpha$ and $\beta$, and (d) characteristic frequency $\omega_1=(G_0/\mathbb{G})^{1/\beta}$. Spectra measured on gels exposed to $\gamma_{\rm OS} < \gamma_c\simeq 20\%$ were fitted by a Fractional Maxwell model [Eq.~\eqref{eq:FMM}, inset in Fig.~\ref{fig:spectra_amplitude}(a)], while those exposed to $\gamma_{\rm OS} >\gamma_c$ were fitted with a Generalized Fractional Kelvin--Voigt model [Eq.~\eqref{eq:GFKVM}, inset in Fig.~\ref{fig:spectra_amplitude}(b)]. The vertical dotted line indicates the critical strain amplitude $\gamma_c\simeq 20\%$ beyond which oscillations imposed throughout gelation for $\mathcal{T}_{\rm OS}=2\,t_g$ significantly impact the long-term linear viscoelastic properties of the gel. Error bars denote 95\% confidence intervals from the fit. In (c), the horizontal gray line marks the fixed value $\alpha=0.43$, imposed to all fits, irrespective of $\gamma_{\rm OS}$. In (c) and (d), red curves serve as a guide for the eye.}
\label{fig:fractionalparameters}
\end{figure}

To further assess the impact of oscillatory forcing applied through gelation on the final gel properties, we now turn to the frequency dependence of the viscoelastic spectra measured at long times ($t=5\,t_g$). Figure~\ref{fig:fig2} summarizes the effect of oscillation duration on gels formed under quiescent conditions and under LAOS at a strain amplitude of $40\%$. In the absence of large-amplitude oscillations [Fig.~\ref{fig:fig2}(a)], the storage modulus $G'$ is nearly frequency independent, while the loss modulus $G''$ decreases monotonically with frequency, consistent with the behavior expected for similar silica gels formed under quiescent conditions \cite{Aime:2018, Keshavarz:2021}. When the gel formation is conducted under LAOS [Fig.~\ref{fig:fig2}(b)] for a duration $\mathcal{T}_{\rm OS}=1.6\,t_g$, the frequency dependence of $G'$ measured at long time remains unchanged. However, its amplitude is significantly reduced, which is consistent with a weakening of the network. More strikingly, the loss modulus no longer displays a simple monotonic decay: for a sufficiently long duration of oscillations applied during gelation, $G''(\omega)$ develops a minimum and exhibits an upturn at high frequencies, signaling the emergence of an additional dissipative process. This behavior is emphasized in Fig.~\ref{fig:fig2}(c), which shows the loss tangent $\tan \delta =G''/G'$ as a function of frequency $\omega$ for different oscillation durations. Although gels formed under quiescent conditions remain predominantly elastic at high frequencies, those subjected to prolonged oscillations display a crossover toward more dissipative behavior. These results demonstrate that LAOS imposed throughout the gelation process not only reduces the gel elasticity but also qualitatively alters the viscoelastic spectrum at high frequency, which suggests the presence of some damage-induced dissipation.

To rationalize these spectral features, we analyze the data within the framework of fractional viscoelastic models \cite{Bonfanti:2020}. These mechanical models are based on a spring-pot element that interpolates between a spring and a dashpot and is defined by a constitutive equation between stress $\sigma$ and strain $\gamma$ that reads $\sigma=\mathbb{V} \mathrm{d}^\alpha \gamma/\mathrm{d}t^\alpha$, where $\alpha$ is a dimensionless exponent ($0 \leq \alpha \leq 1$) and $\mathbb V$ is a quasi-property with dimension $\rm Pa\ s^\alpha$ \cite{Jaishankar:2013}. When formed under quiescent conditions, the present silica gel is accurately described by a Fractional Maxwell (FM) model composed of a spring in series with a spring-pot element [see sketch in Fig.~\ref{fig:fig2}(a)], which captures both the weak frequency dependence of $G'$ and the power-law decay of $G''$, as first shown in Ref.~\cite{Keshavarz:2021}. The FM model reads
\begin{equation}
    G^*(\omega)=G_0\frac{(i\omega/\omega_0)^\alpha}{1+ (i\omega/\omega_0)^\alpha} \, , \label{eq:FMM}
\end{equation}
with $\omega_0=(G_0/\mathbb{V})^{1/\alpha}$ the characteristic frequency scale that sets the boundary between a spring-dominated regime for $\omega\gg \omega_0$ and a spring-pot dominated regime for $\omega\ll \omega_0$. The fit of the data yields the following parameters (including 95\% confidence intervals): $\omega_0=(7.9\pm 0.5)~\times 10^{-4}~\rm rad\ s^{-1}$, $\alpha=0.43 \pm 0.02$, and $G_0=1090\pm20~\rm Pa$. Note that all data are described satisfactorily by fixing $\alpha =0.43$, which ensured better robustness in the fitting procedure for all models employed thereafter.

In contrast, gels formed under LAOS require a different model to capture 
the upturn of $G''$ at high frequencies. Here, we use a Generalized Fractional Kelvin--Voigt (GFKV) model composed of a spring ($G_0$) and a spring-pot ($\mathbb{G}$, $\beta$) in parallel, placed in series with a second spring-pot ($\mathbb{V}$, $\alpha$) [see the sketch in Fig.~\ref{fig:fig2}(b)]. 
The GFKV model reads
\begin{equation}
    G^\ast(\omega) = G_0(i\omega/\omega_0)^\alpha\frac{1+ (i\omega/\omega_1)^\beta }{1+(i\omega/\omega_0)^\alpha+(i\omega/\omega_1)^\beta} \, , \label{eq:GFKVM}
\end{equation}
where $\omega_0=(G_0/\mathbb{V})^{1/\alpha}$ and $\omega_1=(G_0/\mathbb{G})^{1/\beta}$ are the two characteristic frequencies built with the spring and each of the spring-pot elements.
The best fit of the data in Fig.~\ref{fig:fig2}(a) yields $\omega_0=(7\pm 1)\times 10^{-3}~\rm rad\ s^{-1}$, $\omega_1=(3.1\pm 0.8)\times 10^{3}~\rm rad\ s^{-1}$, and $\beta=0.5\pm0.1$, with $\alpha$ set to $0.43$.

To identify a systematic trend, we have prepared seven gels exposed to $\gamma_{\rm OS}=40\%$ for durations $\mathcal{T}_{\rm OS}$ varying between $0$ (for the gel formed under quiescent conditions) and $2.1\,t_g$. The viscoelastic spectrum of each gel is measured at long time ($t=5\,t_g$) and fitted by the FM model for $\mathcal{T}_{\rm OS}<\mathcal{T}_c$ and by the GFKV model for $\mathcal{T}_{\rm OS}\geq\mathcal{T}_c$. The corresponding fitting parameters are reported in Fig.~\ref{fig:40pctFitParams}. Let us first examine the parameters $G_0$, $\alpha$, and $\omega_0$, which are common to both models. One can see that the elastic modulus $G_0$ decreases sharply for $\mathcal{T}_{\rm OS}\geq\mathcal{T}_c$ in agreement with the measurement of $G'$ conducted at $\omega=2\pi~\rm rad \ s^{-1}$, and reported in the inset of Fig.~\ref{fig:evolutionLAOS}. Strikingly, the exponent $\alpha$ can be set to the same constant value $\alpha =0.43$ irrespective of $\mathcal{T}_{\rm OS}$, which means that $\alpha$ corresponds to a gel characteristic that is not affected by the extended period spent under LAOS. In contrast, the frequency $\omega_1$ and the exponent $\beta$ associated with the GFKV model only emerge for $\mathcal{T}_{\rm OS}> \mathcal{T}_c$: $\omega_1$ displays a sharp decrease for increasing oscillatory duration $\mathcal{T}_{\rm OS}$, while $\beta$ increases from 0.2 to about 0.7.

The need for an additional fractional element highlights the presence of a new dissipation channel, consistent with the upturn observed in $G''(\omega)$ for $\mathcal{T}_{\rm OS}> \mathcal{T}_c$. This change of model thus provides strong --yet indirect-- evidence that large-amplitude oscillatory forcing conducted during gelation alters the gel microstructure, introducing damage-induced relaxation modes absent under quiescently formed gels. The decreasing trend observed in $\omega_1$ suggests that such modes involve larger and larger length scales as $\mathcal{T}_{\rm OS}$ increases, and the fact that $\beta$ gets closer and closer to 1, where the additional spring-pot would reduce to a simple viscous dashpot, hints at predominantly viscous dissipation modes.

\begin{figure*} 
\centering
\includegraphics[width=\textwidth]{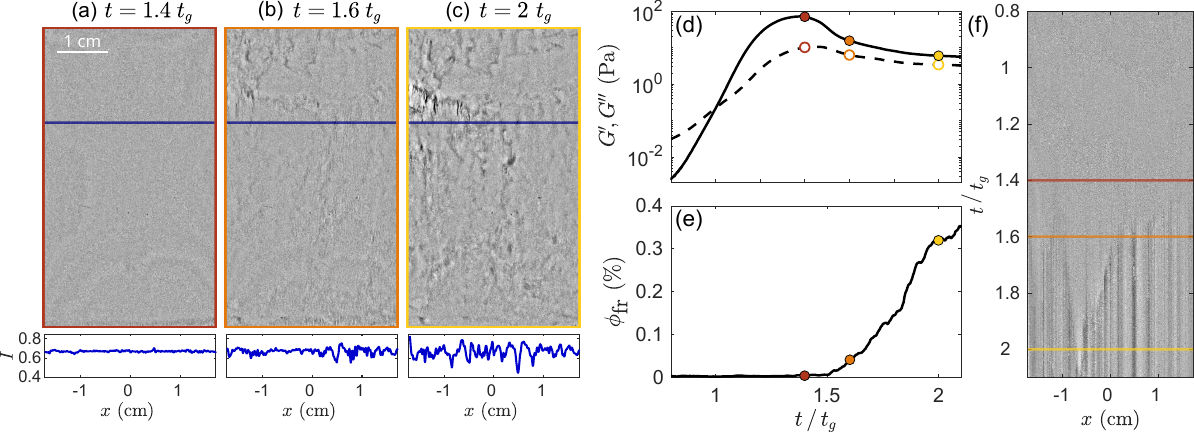}
\caption{Direct visualization of gel formation under LAOS with amplitude $\gamma_{\rm OS}=40\%$ in a transparent Couette cell. Panels (a)--(c) show background-subtracted images obtained by subtracting the image at $t=0$ from each raw frame recorded at (a)~$t=1.4\,t_g$, (b)~$t=1.6\,t_g$, and (c)~$t=2\,t_g$, with the corresponding horizontal intensity profile (blue lines) shown below each image. (d) Elastic modulus $G'$ (solid line) and viscous modulus $G''$ (dashed line) at $\omega=2\pi~\rm rad\ s^{-1}$ as a function of normalized time $t/t_g$. Colored markers correspond to the times shown in (a-c). (e)~Fraction of pixels, $\phi_{\rm fr}$, whose intensity falls below a threshold -- defined as the lowest pixel intensity measured in the image at $t=t_g$ -- plotted as a function of $t/t_g$. (f)~Spatiotemporal diagram of the intensity profile along the blue line in (a)--(c), highlighting the onset and growth of fracture patterns. See Movie~1 in the Supplemental Material \cite{SM} for a dynamic version of this figure.}
\label{fig:video40pc}
\end{figure*}

\subsection{Impact of oscillation amplitude on gel elastic properties}
\label{sec:amplitude}

Having established that, beyond a critical duration, large-amplitude oscillations affect the linear viscoelastic properties of silica gels, we now turn to the role of the strain amplitude. To this end, we fix the oscillation duration to $\mathcal{T}_{\rm OS}=2\,t_g$ and vary $\gamma_{\rm OS}$ between $1\%$ and $80\%$. During gelation, all experiments exhibit a similar early-time increase of $G'$, followed by a maximum whose height and position in time $t_\mathrm{max}$ depend on $\gamma_{\rm OS}$ (see Fig.~\ref{fig:influenceAmp} in Appendix~\ref{appendix:strainamp}). 
Notably, similar to the effect of increasing the strain amplitude, an increase in the oscillation frequency also shifts this maximum towards shorter reduced times $t_\mathrm{max}/t_g-1$, as illustrated in Fig.~\ref{fig:influenceFreq} in Appendix~\ref{Appendix:frequency}. This indicates that both larger deformations and more frequent disruptions increasingly perturb the nascent network as it forms, accelerating its response to the imposed oscillations.
We next examine how this amplitude-dependent mechanical history influences the long-time viscoelastic properties of the gels.

Figure~\ref{fig:spectra_amplitude} shows the viscoelastic spectra of the silica gels measured by SAOS at long times after gelation ($t=5\,t_g$). For each experiment, a freshly prepared sample was subjected to oscillatory shear of amplitude $\gamma_{\rm OS}$ for a fixed duration $\mathcal{T}_{\rm OS}=2\,t_g$, after which the linear viscoelastic spectra were recorded (at $\gamma_0=0.5\%$). The applied strain amplitude was systematically varied between $\gamma_{\rm OS}=1\%$ and $80\%$. Figures~\ref{fig:spectra_amplitude}(a) and \ref{fig:spectra_amplitude}(b) illustrate two representative cases, after SAOS at $\gamma_{\rm OS}=10\%$ and after LAOS at $\gamma_{\rm OS}=40\%$), respectively. Figure~\ref{fig:spectra_amplitude}(c) shows the loss tangent, $\tan \delta = G''/G'$, as a function of frequency for all tested strain amplitudes, ranging from $\gamma_{\rm OS}=1\%$ to $80\%$ (dark to bright colors). For $\gamma_{\rm OS} < \gamma_c \simeq 20\%$, the spectra are well described by the Fractional Maxwell model introduced in Sect.~\ref{sec:duration} to fit the spectrum formed under SAOS [see red curves in Fig.~\ref{fig:spectra_amplitude}(a) and \ref{fig:spectra_amplitude}(c)]. In contrast, for $\gamma_{\rm OS} > \gamma_c$, the frequency dependence of $\tan \delta$ deviates from a simple Fractional Maxwell model, while solid-like properties are preserved, as evidenced by the finite low-frequency elastic modulus. In particular, $\tan \delta$ shows an upturn at high frequencies, which is all the more pronounced as the strain amplitude is large. In this large-amplitude regime, the Generalized Fractional Kelvin--Voigt model introduced in Sect.~\ref{sec:duration} successfully fits the long-time viscoelastic spectra [see red curves in Fig.~\ref{fig:spectra_amplitude}(b) and \ref{fig:spectra_amplitude}(c)].
 
Figure~\ref{fig:fractionalparameters} summarizes the parameters extracted from fitting the linear viscoelastic spectra with the Fractional Maxwell for $\gamma_{\rm OS} < \gamma_c$ and with the Generalized Fractional Kelvin--Voigt model for $\gamma_{\rm OS} > \gamma_c$. The elastic modulus $G_0$ remains nearly constant, up to $\gamma_c$, beyond which it decreases sharply, indicating a softening of the gel network. The opposite trend is observed in the characteristic frequency $\omega_0$, which increases markedly with increasing $\gamma_{\rm OS}$ beyond $\gamma_c$. 
By contrast, the fractional exponent $\alpha$ [Fig.~\ref{fig:fractionalparameters}(c), circles] remains nearly constant and can be set to $\alpha=0.43$ across all strain amplitudes, indicating that the broad distribution of relaxation modes encoded in this parameter is preserved. In turn, the exponent $\beta$ introduced for $\gamma_{\rm OS} >\gamma_c$ [Fig.~\ref{fig:fractionalparameters}(c), squares] increases markedly with increasing $\gamma_{\rm OS}$, suggesting that gels prepared under stronger oscillatory forcing exhibit a more viscous-like dissipative response. Finally, the additional frequency scale $\omega_1$ [Fig.~\ref{fig:fractionalparameters}(d)] decreases systematically with $\gamma_{\rm OS}$, consistent with the emergence of a slower relaxation process in gels formed at high strain amplitudes.

\subsection{Rheo-optical measurements and evidence for crack growth under LAOS}
\label{sec:rheoOptics}

In order to further explore the changes induced by LAOS during gelation, we performed simultaneous rheology at $\gamma_{\rm OS}=40\%$ and imaging experiments in a transparent Couette cell (see Sect.~\ref{sec:rheology} for technical details). Figures~\ref{fig:video40pc}(a)–(c) display representative images at three distinct times after the overshoot of $G'$ [Fig.~\ref{fig:video40pc}(d)]. The same data are shown in Movie~1 in the Supplemental Material \cite{SM}. At early times, the gel appears homogeneous [Fig.~\ref{fig:video40pc}(a) for $t=1.4t_g$]. Past the critical time $\mathcal{T}_c=1.45\,t_g$ identified in Fig.~\ref{fig:evolutionLAOS}, the texture becomes macroscopically heterogeneous [see Fig.~\ref{fig:video40pc}(b) for $t = 1.5\,t_g$], and large-scale fractures develop across the field of view [see Fig.~\ref{fig:video40pc}(c) for $t = 1.8\,t_g$].

Figure~\ref{fig:video40pc}(e) shows that the fraction of pixels below the minimum intensity over the entire image at $t=t_g$ increases sharply for $t > \mathcal{T}_c$, indicating that the nucleation and growth of fracture-like patterns coincide with the decrease in the gel elastic properties reported in Fig.~\ref{fig:evolutionLAOS} for $t \geq \mathcal{T}_c$ [see also Fig.~\ref{fig:video40pc}(f) for a spatiotemporal diagram of the intensity profiles at a given height in the Couette cell]. Together, these observations establish a direct correlation between the rheological softening of the gel beyond the overshoot of $G'$ and the onset of fracturing at scales typically larger than $3~\mu$m, the size of the pixel in Figs.~\ref{fig:video40pc}(a)–(c).

\subsection{Intracycle analysis of LAOS waveforms}
\label{sec:Intracycle}

\begin{figure*} 
\centering
\includegraphics[width=\textwidth]{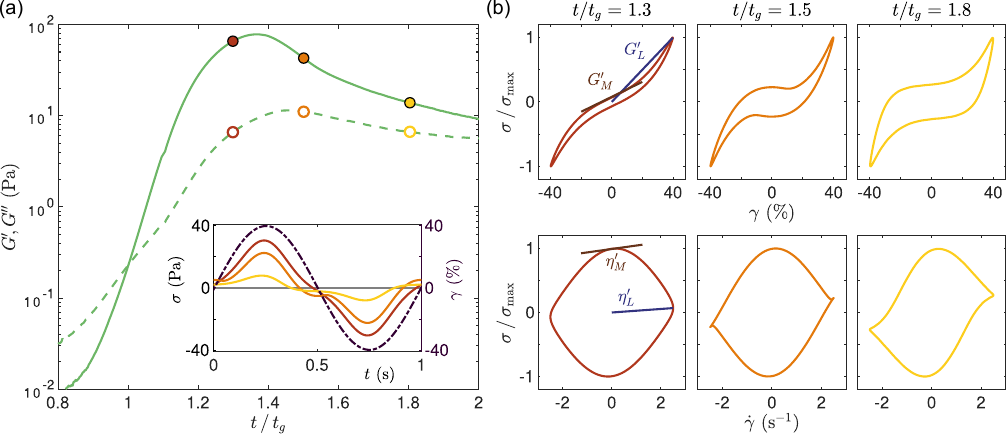}
\caption{Rheological response of the silica suspension during gelation under LAOS at $\gamma_{\rm OS}=40\%$. (a) Elastic modulus $G'$ (solid line) and viscous modulus $G''$ (dashed line) as a function of normalized time $t/t_g$. Inset: stress response (solid lines, left axis) to the imposed strain (dash-dotted line, right axis), with colors corresponding to the symbols in the main plot.
(b) Lissajous–-Bowditch representations of particular oscillation cycles at times $t/t_g=1.3, 1.5,$ and $ 1.8$ plotted as stress--strain curves (``elastic'' representation, top row) and stress--shear-rate curves (``viscous'' representation, bottom row). The stress is normalized by its maximum value $\sigma_\mathrm{max}$ over the whole cycle. The purple/blue solid lines indicate the slopes used to estimate the following intracycle moduli and viscosities: the minimum-strain modulus $G'_M = \left.{\mathrm{d}\sigma}/{\mathrm{d}\gamma}\right|_{\gamma=0}$, the large-strain modulus $G'_L = \left.{\sigma}/{\gamma}\right|_{\gamma=\gamma_{\rm OS}}$, the minimum-rate viscosity $\eta'_M = \left.{\mathrm{d}\sigma}/{\mathrm{d}\dot{\gamma}}\right|_{\dot{\gamma}=0}$, and the large-rate viscosity $\eta'_L = \left.{\sigma}/{\dot{\gamma}}\right|_{\dot{\gamma}=\dot{\gamma}_{\rm OS}}$, with $\dot \gamma_{\rm OS}=\omega \gamma_{\rm OS}$. See Movie~2 in the Supplemental Material \cite{SM} for a dynamic version of this figure. 
}
\label{fig:Lissajous}
\end{figure*}

In this section, we investigate the nonlinear rheological response of the silica suspension during the sol-gel transition by complementing conventional oscillatory tests with intracycle analysis of LAOS data acquired thanks to the strain-controlled ARES-G2 rheometer. Figure~\ref{fig:Lissajous}(a) shows the time evolution of the viscoelastic moduli under oscillatory shear at $\gamma_{\rm OS}=40\%$, reproducing the characteristic overshoot of $G'$ followed by a slow decay, as previously shown in Fig.~\ref{fig:evolutionLAOS}. The inset highlights the strongly distorted stress waveforms, which progressively deviate from a sinusoidal shape as gelation proceeds, reflecting the increasing nonlinearity of the response. 
This behavior is further quantified in Fig.~\ref{fig:Lissajous}(b) by Lissajous--Bowditch plots at different stages of the gelation process in both the ``elastic'' representation, i.e., stress $\sigma(t)$ vs. strain $\gamma(t)$ (top row) and the ``viscous'' representation, i.e., stress $\sigma(t)$ vs. shear-rate $\dot\gamma(t)$ (bottom row) \cite{Ewoldt:2008,Ewoldt:2010}. At early times ($t=1.3\,t_g$), the Lissajous–Bowditch curves already display appreciable anharmonic features, signaling the early development of intracycle nonlinearity prior to the overshoot. 
At later times ($t=1.8\,t_g$), the loops exhibit stronger and stronger deviations from linearity, with persistent intracycle stiffening. Overall, the intracycle analysis indicates that the overshoot in $G'$ coincides primarily with the build-up of elastic nonlinearity, while the form of the intracycle viscous response remains largely unchanged.

\begin{figure} 
\centering
\includegraphics[width=8.6 cm]{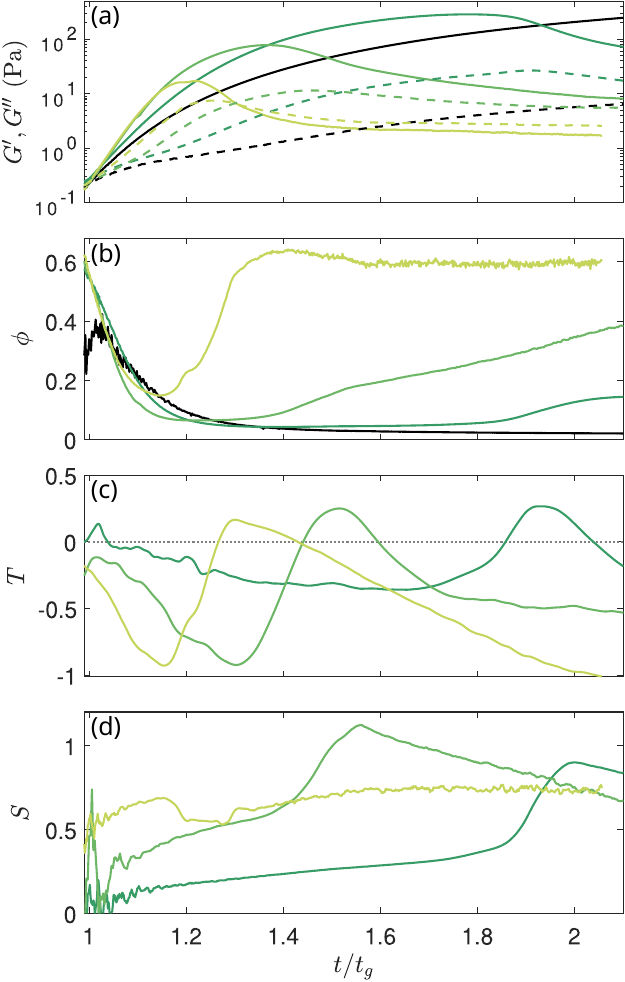}
\caption{Intracycle analysis of gelation under LAOS at $\gamma_{\rm OS}=25$, $40$, and $80\%$ from dark green to light green. (a) Elastic modulus $G'$ (solid lines) and viscous modulus $G''$ (dashed lines) measured at $\omega=2\pi~\rm rad\ s^{-1}$, (b) energy dissipation ratio $\phi$, (c) intracycle shear-thickening parameter $T$, and (d) intracycle strain-stiffening parameter $S$ as a function of normalized time $t/t_g$. Black lines in (a) and (b) correspond to the gelation under SAOS at $\gamma_{\rm OS}=1\%$.}
\label{fig:intracycle}
\end{figure}

Having established the nature of intracycle nonlinearities at a representative strain amplitude of $\gamma_{\rm OS}=40\%$, we vary the $\gamma_{\rm OS}$ across the LAOS regime to strongly nonlinear conditions and report on both conventional viscoelastic moduli and intracycle descriptors. Figure~\ref{fig:intracycle}(a) shows that the strain amplitude modifies the build-up and subsequent evolution of $G'$ and $G''$. For all amplitudes under study, $\gamma_{\rm OS}=25\%$, $40\%$, and $80\%$, $G'$ rises steeply and crosses $G''$, then displays a pronounced overshoot before settling into a regime with $G' > G''$ for $\gamma_{\rm OS}=25\%$ and $40\%$. At the largest amplitude $\gamma_{\rm OS}=80\%$, an overshoot of $G'$ is still observed, but the post-overshoot decay leads to $G' \lesssim G''$. Thus, increasing the driving amplitude weakens the long-time elastic properties and alters the rheological signature of gelation without fully suppressing the transition. 
Figure~\ref{fig:intracycle}(b) shows the evolution of the energy dissipation ratio $\phi$, which compares the energy dissipated in a single LAOS cycle to the energy which would be dissipated in a rigid, perfect plastic response with equivalent strain amplitude $\gamma_{\rm OS}$, and maximum stress $\sigma_{\rm max}$. In practice, $\phi= \pi G' \gamma_{\rm OS}/(4\sigma_{\rm max})$ \textcolor{black}{with $\phi=0$ for a purely elastic response and $\phi=\pi/4$ for Newtonian behavior, while $\phi=1$ corresponds to the ideal perfectly plastic limit and serves only as a reference value} \cite{Ewoldt:2010}. At gelation time $t = t_g$, we observe that $\phi \simeq 0.6$, a value close to the Newtonian case. For all strain amplitudes, $\phi$ then decreases for $t/t_g\gtrsim 1$, marking the emergence of an increasingly elastic gel network. Whilst gels formed in quiescent conditions display a monotonic decrease in $\phi$, gels formed under LAOS display a minimum in $\phi$ followed by an increase that coincides with the overshoot in $G''$. This behavior indicates that higher $\gamma_{\rm OS}$ enhances intracycle energy dissipation even as the sample remains predominantly solid-like, weakening the long-time elastic properties in agreement with the results of Sect.~\ref{sec:amplitude}.

The intracycle parameters $T$ and $S$ measured during gelation under LAOS are displayed in Figs.~\ref{fig:intracycle}(c) and (d), respectively. The shear-thickening parameter $T$ follows a complex yet reproducible evolution, decreasing from values close to 0 at early time, which is indicative of a stronger intracycle shear-thinning, and going through a minimum that corresponds to the minimum in $\tan\delta(t)$ reached slightly before the overshoot $G'(t)$. $T$ then increases, reaching slightly positive values indicative of weak intracycle shear-thickening, and goes through a maximum that corresponds to the inflection point in $G'(t)$ [see Fig.~\ref{fig:intracycle}(c)]. On the other hand, the strain-stiffening parameter $S$ remains always positive, indicating significant intracycle strain-stiffening throughout gelation under LAOS. For $\gamma_{\rm OS}=25\%$ and $40\%$, $S$ starts to increase markedly at the $G'$ overshoot and goes through a maximum that is concomitant to the one in $T$, i.e., to the inflection point in $G'(t)$. However, for $\gamma_{\rm OS}=80\%$, $S$ only shows a small drop after the overshoot in $G'(t)$ [see Fig.~\ref{fig:intracycle}(d)]. 

\begin{figure*} 
\centering
    \includegraphics[width=\textwidth]{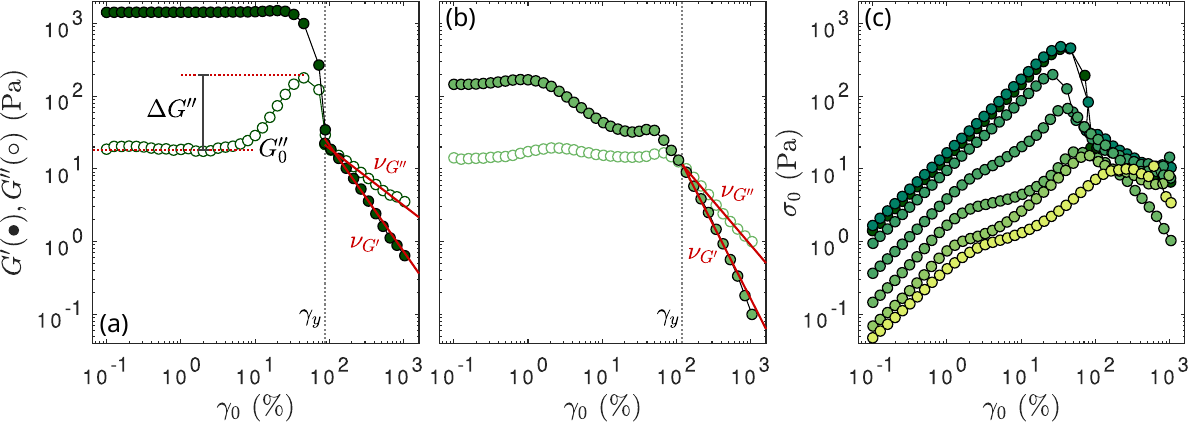}
\caption{Elastic modulus $G'$ and viscous modulus $G''$ measured at $\omega=2\pi~\rm rad\ s^{-1}$ upon increasing the strain amplitude $\gamma_0$ for silica gels probed at long times ($t=5\,t_g$) and previously formed (a) under SAOS ($\gamma_{\rm OS}=1\%$) and (b) under LAOS with amplitude $\gamma_{\rm OS}=40\%$. The oscillatory shear at $\gamma_{\rm OS}$ is applied for a duration of $\mathcal{T_{\rm OS}}=2\,t_g$ before switching to a small amplitude $\gamma_0=0.5\%$ for another $3\,t_g$. (c) Stress response $\sigma_0=G' \gamma_0$ for silica gels formed under oscillatory shear with amplitude $\gamma_{\rm OS}=1,\ 10,\ 25,\ 30,\ 40,\ 50$, and $80\%$ (from dark to bright) applied for a duration $\mathcal{T_{\rm OS}}=2\,t_g$. Color code consistent with that used in Figs.~\ref{fig:spectra_amplitude}, \ref{fig:fractionalparameters}, \ref{fig:Lissajous}(a), and \ref{fig:intracycle}.}
\label{fig:gammaOS_sweep}
\end{figure*}

In summary, both viscous and elastic intracycle signatures seem to be tightly coupled to the transient structural event that produces the overshoot in $G'(t)$.
The peak and timing of the overshoot set the clock for the shear-thickening parameter $T$, which undergoes a reproducible sign change immediately after the overshoot. This result suggests that, within a cycle of oscillation, the dominant dissipative mechanism corresponds to the viscous dissipation within the cracks. Meanwhile, the strain-stiffening parameter $S$ remains positive at all times, suggesting that intracycle strain-stiffening may be the hallmark of gelation conducted under large-amplitude oscillations. 

\subsection{Nonlinear response of silica gels formed under different strain amplitudes}
\label{sec:Strainsweep}

In this final section, we examine the nonlinear response of the silica gels formed under oscillatory shear at various strain amplitudes. In practice, silica gels are formed under oscillatory shear at a strain amplitude $ \gamma_{\rm OS}$ ranging from $1\%$ to $80\%$, applied for $t=2\,t_g$, before monitoring the subsequent aging of the gel under SAOS, at $\gamma_0=0.5\%$. When the sample reaches age $t=5\,t_g$, we perform the amplitude of the strain is swept logarithmically from $\gamma_0=0.1\%$ to $1000\%$, hereafter referred to as ``LAOStrain test.'' 

The LAOStrain response of a silica gel formed under SAOS at $\gamma_{\rm OS}=1\%$ is shown in Fig.~\ref{fig:gammaOS_sweep}(a). Both $G'$ and $G''$ remain constant in the linear regime, which extends up to $\gamma_0 \simeq 10\%$. Beyond this point, $G''$ exhibits a pronounced overshoot with a peak at $\gamma_0 \simeq 50\%$, while $G'$ exhibits a sharp drop by two decades for $\gamma_0 \geq 30\%$. The yield point $\gamma_y$, defined as the cross-over of $G'$ and $G''$, occurs at $\gamma_0 \simeq 80\%$, beyond which $G'$ and $G''$ follow a power-law decay, i.e., $G' \propto \gamma_0^{-\nu_{G'}}$ and $G'' \propto \gamma_0^{-\nu_{G''}}$, with $\nu_{G'}=1.55\pm0.04$ and $\nu_{G''}=0.79\pm0.02$. Such a LAOStrain behavior corresponds to a ``type III'' yielding scenario, as classified in the literature \cite{Hyun:2002,Hyun:2011}.
For comparison, we show the LAOStrain response of a silica gel formed under LAOS at $\gamma_{\rm OS}=40\%$ in Fig.~\ref{fig:gammaOS_sweep}(b). In this case, the low-strain plateau of $G'$ and $G''$ is reduced compared to the gel formed under SAOS, in agreement with the linear viscoelastic measurements reported in Sect.~\ref{sec:amplitude}. The elastic modulus exhibits a two-step decrease, with a first drop starting at $\gamma_0 \simeq 2\%$, and a second one at about $\gamma_0 \simeq 50\%$. Such drops in $G'$ are concomitant with slight overshoots in $G''$. Finally, $G'$ and $G''$ cross over at $\gamma_y \simeq 100\%$, beyond which $G'$ and $G''$ follow a power-law decay, $G' \propto \gamma_0^{-\nu_{G'}}$ and $G'' \propto \gamma_0^{-\nu_{G''}}$, with $\nu_{G'}=2.0\pm0.1$ and $\nu_{G''}=1.20\pm0.03$.

Figure~\ref{fig:gammaOS_sweep}(c) gathers the LAOStrain responses for $\gamma_{\rm OS}=1\%$ to $80\%$ in terms of the stress amplitude $\sigma_0=G' \gamma_0$ plotted as a function of the strain amplitude $\gamma_0$. It shows that the two trends respectively observed in Figs.~\ref{fig:gammaOS_sweep}(a) and (b) occur on each side of the critical amplitude $\gamma_c$ already identified in Sect.~\ref{sec:amplitude}. Indeed, gels prepared under low-amplitude oscillations, $\gamma_{\rm OS} < \gamma_c \simeq 20\%$, display a brittle-like response, where the stress drops abruptly following the end of the linear regime. In contrast, for $\gamma_{\rm OS} > \gamma_c$, strong strain-softening is observed for $\gamma_0 \gtrsim 1\%$, and the subsequent stress overshoot is of a lesser amplitude, and occurs at a greater strain amplitude than for $\gamma_{\rm OS} < \gamma_c$. Last but not least, the stress also shows a much smoother decay beyond the maximum, pointing to a more ductile-like response \cite{Divoux:2024}.

\begin{figure} 
\centering
\includegraphics[width=8.6 cm]{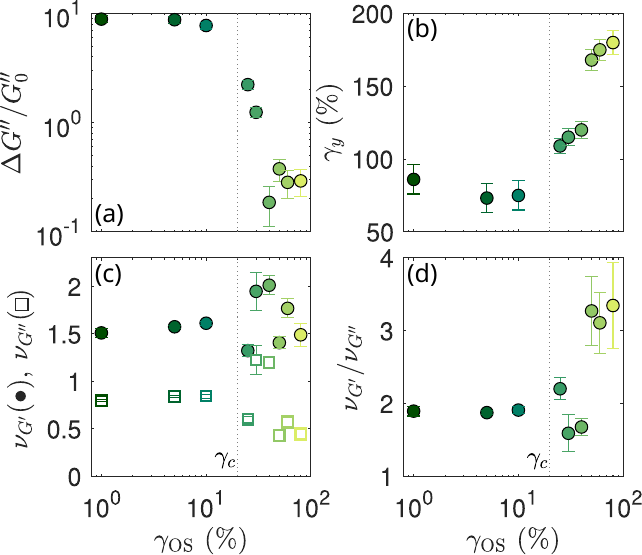}
\caption{Characteristics of the LAOStrain response of gels formed under oscillatory shear at various amplitudes $\gamma_{\rm OS}$. (a)~Relative amplitude of the peak in $G''$ measured close to the yield point, defined as the crossover between $G'$ and $G''$. (b) Yield strain $\gamma_y$ defined by the crossover of $G'$ and $G''$. (c)~Power-law exponent $\nu_{G'}$ and $\nu_{G''}$ characterizing the decay of $G'$ and $G''$ beyond the yield point. (d) Ratio $\nu_{G'}/\nu_{G''}$. The vertical dotted lines highlight the critical strain $\gamma_c$ identified in Sect.~\ref{sec:amplitude}.}
\label{fig:gammaOSamplitudeParam}
\end{figure}

More quantitatively, Fig.~\ref{fig:gammaOSamplitudeParam}(a) shows that the relative amplitude of the overshoot in $G''$ near yielding remains essentially constant for gels formed under oscillatory shear with $\gamma_{\rm OS} < \gamma_c$, but decreases sharply for $\gamma_{\rm OS} > \gamma_c$, in line with the progressive brittle-to-ductile transition visible in Fig.~\ref{fig:gammaOS_sweep}(c) \cite{Kamani:2024}. In parallel, Fig.~\ref{fig:gammaOSamplitudeParam}(b) shows that the apparent yield strain $\gamma_y$, defined as the crossover of $G'$ and $G''$, remains constant at $\gamma_y \simeq 80\%$ for $\gamma_{\rm OS} < \gamma_c$ but increases significantly for $\gamma_{\rm OS} > \gamma_c$. Beyond the yield point, the power-law exponents extracted from the decay of $G'$ and $G''$ also evolve systematically with the oscillation amplitude applied during gelation [Fig.~\ref{fig:gammaOSamplitudeParam}(c)]. For $\gamma_{\rm OS} < \gamma_c$, both exponents remain constant, and their ratio $\nu_{G'}/\nu_{G''} \simeq 2$, which is the typical value for a soft solid displaying a Maxwell-like response in the vicinity of the yield point \cite{Miyazaki:2006,Wyss:2007}. In contrast, for $\gamma_{\rm OS} > \gamma_c$, $\nu_{G'}$ and $\nu_{G''}$ show a larger dispersion. For $\gamma_{\rm OS}\gtrsim 50\%$, the ratio $\nu_{G'}/\nu_{G''}$ increases from about 2 to more than 3 [Fig.~\ref{fig:gammaOSamplitudeParam}(d)], indicating that gels formed under large-amplitude oscillations depart from the simple Maxwell-like behavior observed near yielding for $\gamma_{\rm OS} < \gamma_c$. This trend is consistent with their distinct linear viscoelastic signature, which evolves from a Fractional Maxwell to a Generalized Fractional Kelvin–Voigt response as $\gamma_{\rm OS}$ increases (see section~\ref{sec:amplitude}). Together, the four LAOStrain observables, $\Delta G''/G''_0$, $\gamma_y$, $\nu_{G'}$, and $\nu_{G''}$, confirm the existence of a critical strain amplitude $\gamma_c \simeq 20\%$, above which the oscillations imposed during gelation significantly alter the subsequent nonlinear mechanical response of the gel at long times. 

\section{Discussion and conclusion}

In this work, we have demonstrated that the application of large-amplitude oscillatory shear during the sol-gel transition of attractive nanoparticles is a powerful method for tuning the final microstructure and mechanical properties of a silica colloidal gel.
We have identified two distinct thresholds: a critical oscillation duration ($\mathcal{T}_c \simeq 1.45\,t_g$, where $t_g$ is the gelation time in the absence of any oscillation) at fixed oscillation amplitude ($\gamma_{\rm OS}=40\%$) and a critical strain amplitude $\gamma_c \simeq 20\%$ at fixed oscillation duration ($\mathcal{T}_{\rm OS}=2\,t_g$). Below these thresholds, oscillatory shear has a modest and reversible effect on gelation. Above them, the gel network is irreversibly damaged, leading to a distinct rheological fingerprint: the elastic modulus $G'$ weakens, while the viscous modulus $G''$ develops a high-frequency increase, signaling a new damage-induced dissipation channel. This transition is elegantly captured by a shift from a simple Fractional Maxwell model for gels formed under small-amplitude oscillations to a Generalized Fractional Kelvin--Voigt model for damaged gels.

Furthermore, our rheo-optical experiments have unambiguously uncovered the physical origin of this rheological shift. The overshoot in $G'$, which marks the onset of irreversible weakening, coincides precisely with the nucleation and propagation of macroscopic cracks throughout the sample. Complementary rheo-SAXS measurements performed at ESRF (see Fig.~\ref{fig:ESRF} in Appendix~\ref{Appendix:ESRF}) show no clear signature of microstructural rearrangements within the probed range of wave vectors, which corresponds to length scales smaller than about $1~\mu\rm m$. 
Such a lack of microscale particle rearrangement contrasts with findings in systems like depletion gels, which exhibit particle-level reorganization under shear \cite{Koumakis:2015}.
This also indicates that the relevant reorganizations in our silica system are either very sparse at the particle scale, or mainly associated with rearrangements in the particle network at scales larger than about $1~\mu\rm m$, consistent with the observed fracture patterns. Therefore, our results provide a direct link between the somewhat abstract parameters of fractional viscoelastic models and a concrete physical process: the additional element in the GFKV model is the mechanical signature of a fractured network.

Such a fractured microstructure also explains the transition observed in the gel nonlinear response for $\gamma_{\rm OS}>\gamma_c$. Indeed, for gels formed under quiescent conditions or under small-amplitude oscillations, the stress increases linearly with the strain amplitude up to the stress maximum, reflecting the uniform deformation of a continuous network. While these gels yield in a brittle-like manner, the LAOS-prepared gels exhibit a more ductile-like yielding. This is most probably because the latter gel network is compartmentalized into domains separated by pre-existing cracks that act as weak planes accommodating strain and preventing the catastrophic stress accumulation that characterizes brittle failure in an intact, homogeneous network. For $\gamma_{\rm OS}>\gamma_c$, the stress would thus depart from linearity at much smaller strain amplitude than for $\gamma_{\rm OS}<\gamma_c$ because part of the imposed strain is accommodated by the opening or sliding of fracture planes rather than by purely elastic deformation. We may therefore interpret the ductile-like yielding of LAOS-prepared gels as a direct consequence of the fracture patterns imprinted during gelation, \textcolor{black}{although the specific effects of oscillation duration and strain amplitude on the fracture process remain to be fully disentangled.}

Beyond the present work, structural and mechanical modifications of a nascent gel network by LAOS appear to be a generic phenomenon across different physical gels, despite differences in the nature of the underlying particle interactions. For instance, studies on whey protein isolate gels subjected to oscillatory shear during heat-induced gelation similarly found that large strains reduce the final elastic modulus and replace the single fracture point of gels formed under quiescent conditions with a two-step yielding associated with greater macroscopic inhomogeneity \cite{Homer:2016}. This effect is also observed in the gelation of agar biopolymer solutions: cooling under quiescent conditions forms a strong, brittle gel that fractures, while cooling under continuous or oscillatory shear produces a weak, cohesive ``soft" gel, demonstrating that shear radically alters the structural development pathway \cite{Altmann:2004,Stokes:2008}. Finally, similar weakening and microstructural modification effects have been reported for methylcellulose hydrogels \cite{Nelson:2022}. This common outcome --converting a strong, brittle gel formed under quiescent conditions into a weak, ductile shear-formed gel-- demonstrates that high strain energy input universally prevents the formation of a pristine, high-modulus network. However, while the mechanism in these biopolymer systems is attributed to microstructural rearrangements or to the enhancement of short-range interactions, our silica system dissipates energy predominantly through the irreversible formation of macroscopic cracks. This specific, macroscopic failure mechanism offers a powerful and direct way to tune both the linear viscoelastic response and the nonlinear yielding behavior.

In order to link macroscopic properties to microscopic phenomena, one may estimate a fracture energy $\Gamma$ using the Griffith rupture criterion \cite{Creton:2016}, $\Gamma \sim \sigma_{\rm f}^{2}L/E$, where $\sigma_{\rm f}$ is the fracture stress, $L$ the flaw length, and $E$ the Young modulus of the material. Taking $\sigma_{\rm f}\simeq 40\;\mathrm{Pa}$, which corresponds to the shear
stress at the maximum of the $G'$ overshoot, $E\simeq G' \simeq 100\;\mathrm{Pa}$, and $L\sim1$~mm, leads to $\Gamma\sim10^{-2}\;\mathrm{J\;m^{-2}}$. When compared to the adhesion energy per unit surface between two colloids, $U/a^2\sim 10^{-3}\;\mathrm{J\;m^{-2}}$, where $U\simeq 10 k_{\rm B}T$ is the attraction energy and $a=6.5$~nm is the particle radius, the fracture energy $\Gamma$ would correspond to the simultaneous rupture of bonds involving about 10 colloids. 
Such rough estimates obviously call for more work to understand how the interplay between interactions at the colloid scale and the oscillatory strain applied during gelation drives the nucleation and growth of macroscopic cracks. Useful insights should be gained both from more detailed rheo-imaging experiments, including, e.g., confocal microscopy or microtomography, and from numerical simulations at the level of individual strands \cite{vanDoorm2018,Verweij2019} or at the level of the strand network \cite{Donley:2022,Bhaumik:2025a,Bhaumik:2025b}.

Looking ahead, the present \textit{tuning-by-damaging} approach opens several intriguing avenues. The observation that higher strain amplitudes lead to more periodic fragments (compare Figs.~\ref{fig:video40pc} and \ref{fig:video80pc}) suggests an underlying shear-induced instability with a characteristic wavelength. A systematic study connecting shear parameters (amplitude, frequency) and geometry (gap size) to this fracture length scale could lead to predictive models for designing materials with controlled heterogeneity \cite{Chaudhury:2015,Fardin:2025}. From a practical standpoint, the imprinting of a controlled fracture pattern into a gel could be a novel strategy for engineering materials with tailored properties, such as enhanced ductility, specific fracture toughness, or targeted energy dissipation, which are desirable for applications in 3D printing or injectable biomaterials. Ultimately, our work shows that shear-induced damage, often seen as an undesirable artifact, can be harnessed as a rational design tool to sculpt the mechanical response of soft solids.

\begin{acknowledgments}
The authors acknowledge the European Synchrotron Radiation Facility (ESRF) for provision of synchrotron radiation facilities under proposal number SC-5099 and thank T.~Narayanan for assistance and support in using beamline ID02. The authors thank G.~Legrand for fruitful discussions and V.~Verdoot for technical support in the use of the ARES-G2 rheometer (TA Instrument). This work benefited from the support of the project ``Fluididense'' ANR-17-CE07-0040 of the French National Research Agency (ANR). 
\end{acknowledgments}

\section*{Data Availibility}
The data corresponding to the figures shown in the article are openly available \cite{DAS}.
\appendix

\section{Determination of the gelation time $t_g$}
\label{sec:GelationTime}

\begin{figure}[h!]
\centering
\includegraphics[width=8.6 cm]{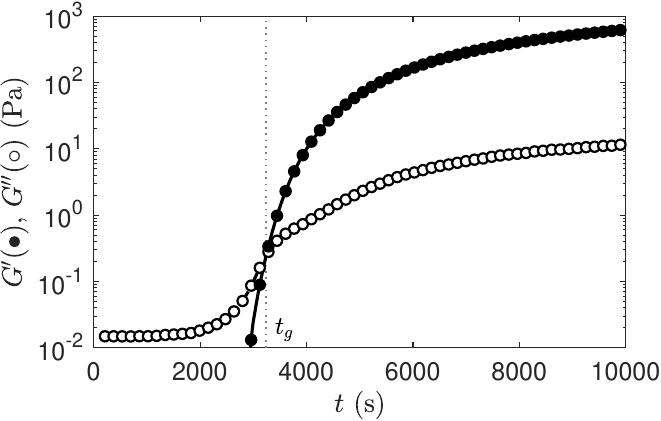}
\caption{Elastic modulus $G'$ and viscous modulus $G''$ as function of time at a fixed probe frequency of $\omega = 2\pi~\rm rad\ s^{-1}$ for an imposed strain amplitude $\gamma_{\rm OS}=1\%$. For this particular experiment, the crossover of $G'$ and $G''$ --used as a proxy for the gelation time $t_g$-- occurs at 3508~s.}
\label{fig:quiescent}
\end{figure}

\begin{figure}[h!]
\centering
\includegraphics[width=8.6 cm]{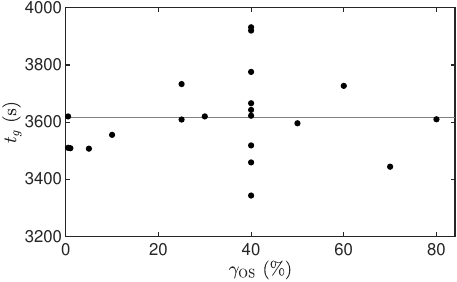}
\caption{The gelation time $t_g$ as determined by the crossover of $G'$ and $G''$ for various imposed strain amplitude $\gamma_{\rm OS}$. The gray horizontal line shows the average $t_g=3590$~s. The standard deviation over all $t_g$ estimates is 190~s.}
\label{fig:tg}
\end{figure}

Figure~\ref{fig:quiescent} shows the quiescent gelation process measured under SAOS. The crossover of $G'$ and $G''$ is used as a proxy to estimate the gelation time $t_g$. Over repeated experiments, we observed $t_g=3590\pm190$~s, independent of $\gamma_{\rm OS}$ (see Fig.~\ref{fig:tg}). In the main text, all times are expressed in units of gelation time $t_g$ determined independently from each experiment by the criterion $G'(t_g)=G''(t_g)$.

\section{Impact of the strain amplitude oscillations on the overshoot in $G'$}
\label{appendix:strainamp}

\begin{figure}[!t] 
\centering
\includegraphics[width=8.6 cm]{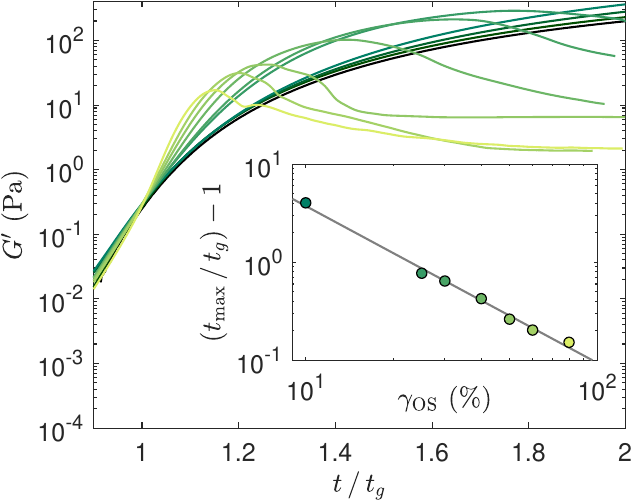}
\caption{Elastic modulus $G'$ as a function of normalized time $t/t_g$ during salt-induced gelation of a colloidal silica suspension under oscillatory shear at $\omega = 2\pi~\rm rad\ s^{-1}$ for various strain amplitudes: $\gamma_{\rm OS}=0.5$, 1, 5, 10, 25, 30, 40, 50, 60, and $80~\%$ (from dark to bright). Inset: time $t_\mathrm{max}$ at which $G'$ reaches its maximum (normalized by $t_g$) as a function of the oscillation amplitude $\gamma_{\rm OS}$. The gray solid line is the best power-law fit of the data, $t_\textrm{max}/t_g - 1\propto\gamma_{\rm OS}^{-1.6}$.}
\label{fig:influenceAmp}
\end{figure}

\begin{figure}[!h] 
\centering
\includegraphics[width=8.6 cm]{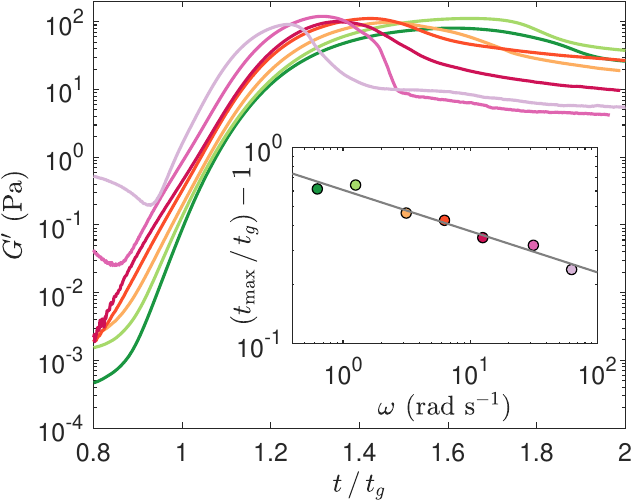}
\caption{Elastic modulus $G'$ as a function of normalized time $t/t_g$ during salt-induced gelation of a colloidal silica suspension under oscillatory shear at $\gamma_{\rm OS}=40\%$ for various oscillation frequencies: $\omega/(2\pi)=0.1$, $0.2$, $0.5$, $1$, $2$, $5$, $10~\rm Hz$ (from green to red to violet). Inset: $t_\mathrm{max}$ as a function of frequency from the main figure. The solid gray line is the best power-law fit of the data, $t_\textrm{max}/t_g-1\propto\omega^{-0.21}$.}
\label{fig:influenceFreq}
\end{figure}

\begin{figure*}[!t] 
\centering
\includegraphics[width=\textwidth]{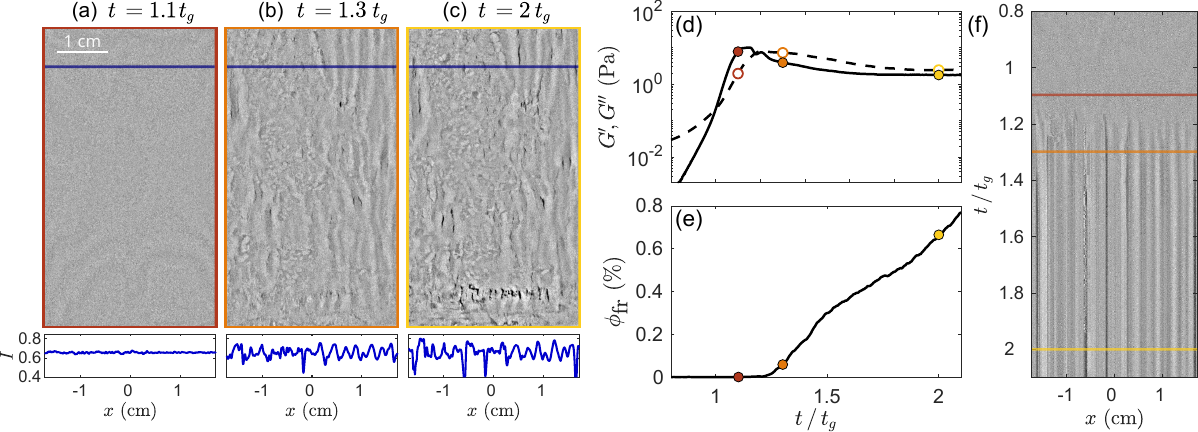}
\caption{Direct visualization of gel formation under LAOS with amplitude $\gamma_{\rm OS}=80\%$ in a transparent Couette cell. Panels (a)--(c) show background-subtracted images obtained by subtracting the image at $t=0$ from each raw frame recorded at (a)~$t=1.1\,t_g$, (b) $t=1.3\,t_g$, and (c) $t=2\,t_g$, with corresponding horizontal intensity profile (blue lines) shown below each image. (d)~Elastic modulus $G'$ (solid line) and viscous modulus $G''$ (dashed line) at $\omega=2\pi~\rm rad\ s^{-1}$ as a function of normalized time $t/t_g$. Colored markers correspond to the times shown in (a-c). (e)~Fraction of pixels, $\phi_{\rm fr}$, whose intensity falls below a threshold -- defined as the lowest pixel intensity measured in the image at $t=t_g$ -- plotted as a function of $t/t_g$. (f)~Spatiotemporal diagram of the intensity profile along the blue line in (a)-(c), highlighting the onset and growth of fracture patterns. See Movie~3 in the Supplemental Material \cite{SM} for a dynamic version of this figure. 
}
\label{fig:video80pc}
\end{figure*}

Figure~\ref{fig:influenceAmp} shows the effect of oscillatory shear amplitude $\gamma_{\rm OS}$ on the evolution of the elastic modulus $G'$ during gelation for gels formed under oscillations of amplitude $\gamma_{\rm OS}$ ranging between $1\%$ and $80\%$ at $\omega=2\pi~\rm rad\ s^{-1}$. While the gelation time $t_g$ remains essentially unchanged, the time $t_{\rm max}$ at which $G'$ reaches a maximum systematically decreases with increasing oscillation amplitude following $t_\textrm{max}/t_g - 1\propto\gamma_{\rm OS}^{-1.6}$. This indicates that network damage occurs earlier under stronger forcing. At large amplitudes, the forming gel experiences partial rupture and rearrangement, which compete with aggregation, leading to the overshoot and subsequent decrease in $G'(t)$ described in the main text. This interpretation is consistent with the rheo-optical observations presented in Sect.~\ref{sec:rheoOptics}, where fracture patterns emerge shortly after the modulus maximum.

\section{Impact of the oscillation frequency on the overshoot in $G'$}
\label{Appendix:frequency}

Figure~\ref{fig:influenceFreq} shows the evolution of the elastic modulus $G'$ during the salt-induced gelation of the Ludox suspension under oscillatory shear at a fixed strain amplitude $\gamma_{\rm OS}= 40\%$ for various oscillation frequencies. The overshoot in $G'(t)$ shifts to earlier times as the oscillation frequency increases. The inset quantifies this trend, showing that the relative time of the overshoot $t_{\rm max}/t_g-1$, decreases weakly with frequency following a power-law dependence $t_{\rm max}/t_g-1 \propto \omega^{-0.21}$.

\section{Rheo-optical measurements in a silica suspension exposed to $\gamma_{\rm OS}=80\%$}

Figure~\ref{fig:video80pc} shows results similar to those reported in Sect.~\ref{sec:rheoOptics} for a gelation performed under oscillatory shear at $\omega=2\pi~\rm rad\ s^{-1}$, but with a larger applied amplitude $\gamma_{\rm OS}=80\%$. The presence of cracks or fractures is first detected at $t/t_g \simeq 1.3$, which corresponds to a lower value than that measured when the gelation is carried out at $\gamma_{\rm OS}=40\%$ (see Fig.~\ref{fig:video40pc} in the main text).

\begin{figure}[!h] 
\centering
\includegraphics[width=8.6 cm]{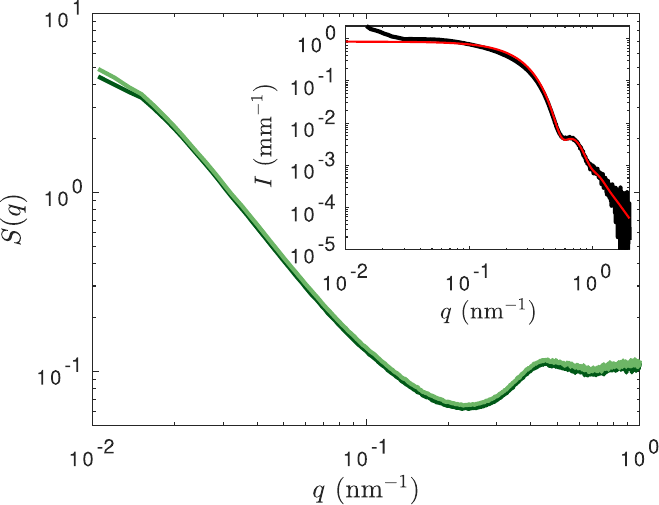}
\caption{Structure factor $S(q)$ of silica gels formed under SAOS ($\gamma_{\rm OS}=3\%$) and LAOS ($\gamma_{\rm OS}=40\%$, applied continuously) measured at $t=2\ t_g$. Both gels display indistinguishable structure factors despite their different viscoelastic spectra. Inset: Form factor $P(q)$ of the silica particles, obtained from a filtered Ludox HS dispersion at 0.1 wt\%. The red curve represents a polydisperse hard-sphere fit with mean particle radius $r_0 = 8.3~\rm{nm}$ and a polydispersity of 13\%.}
\label{fig:ESRF}
\end{figure}

\section{Structural comparison of gels formed under SAOS and LAOS}
\label{Appendix:ESRF}

To compare the microstructure of silica gels formed under SAOS and LAOS, we performed rheometry coupled with Small-Angle X-ray Scattering (SAXS) at the ID02 beamline of ESRF (Grenoble, France). These rheo-SAXS experiments were carried out at room temperature ($\sim$23$^\circ$C) using a stress-controlled rheometer (RS 6000, HAAKE) equipped with a polycarbonate Couette cell of inner radius $R_i=10~\rm mm$, outer radius $R_o=11~\rm mm$, and rotor height $60~\rm mm$. The X-ray beam was aligned in the ``radial" configuration, passing across the Couette cell through the rotation axis of the inner cylinder. Scattering patterns were therefore recorded in the (velocity $\vec{v}$, vorticity $\vec{\nabla} \times \vec{v}$) plane. 
The X-ray beam was monochromatic and highly collimated with a wavelength of $1.013$~\AA. The experiments were conducted at a sample-to-detector distance of $3~\rm m$, covering a scattering vector range of $10^{-2} \leq q \leq 1~\rm nm^{-1}$. Two-dimensional SAXS patterns were recorded with an Eiger2-4M pixel array detector and normalized to an absolute scale after appropriate background subtraction, as well as thickness and transmission normalization (see Ref.~\cite{Narayanan:2022} for technical details).

Figure~\ref{fig:ESRF} shows the structure factor $S(q)$ measured at $t=2\ t_g$ for two different silica gels: one formed under SAOS ($\gamma_{\rm OS}=3\%$) and one formed under LAOS ($\gamma_{\rm OS}=40\%$) with $\mathcal{T}_{\rm OS}=2\ t_g$. The structure factor was calculated by normalizing the scattering intensity $I(q)$ by the particle form factor $P(q)$, shown as an inset in Fig.~\ref{fig:ESRF} and independently measured on a dilute suspension of silica particles. Despite exhibiting radically different viscoelastic spectra, the two gels display indistinguishable structure factors, which confirms that their microstructures are similar on length scales up to $600~\rm nm$.


\begin{thebibliography}{75}%
\makeatletter
\providecommand \@ifxundefined [1]{%
 \@ifx{#1\undefined}
}%
\providecommand \@ifnum [1]{%
 \ifnum #1\expandafter \@firstoftwo
 \else \expandafter \@secondoftwo
 \fi
}%
\providecommand \@ifx [1]{%
 \ifx #1\expandafter \@firstoftwo
 \else \expandafter \@secondoftwo
 \fi
}%
\providecommand \natexlab [1]{#1}%
\providecommand \enquote  [1]{``#1''}%
\providecommand \bibnamefont  [1]{#1}%
\providecommand \bibfnamefont [1]{#1}%
\providecommand \citenamefont [1]{#1}%
\providecommand \href@noop [0]{\@secondoftwo}%
\providecommand \href [0]{\begingroup \@sanitize@url \@href}%
\providecommand \@href[1]{\@@startlink{#1}\@@href}%
\providecommand \@@href[1]{\endgroup#1\@@endlink}%
\providecommand \@sanitize@url [0]{\catcode `\\12\catcode `\$12\catcode
  `\&12\catcode `\#12\catcode `\^12\catcode `\_12\catcode `\%12\relax}%
\providecommand \@@startlink[1]{}%
\providecommand \@@endlink[0]{}%
\providecommand \url  [0]{\begingroup\@sanitize@url \@url }%
\providecommand \@url [1]{\endgroup\@href {#1}{\urlprefix }}%
\providecommand \urlprefix  [0]{URL }%
\providecommand \Eprint [0]{\href }%
\providecommand \doibase [0]{https://doi.org/}%
\providecommand \selectlanguage [0]{\@gobble}%
\providecommand \bibinfo  [0]{\@secondoftwo}%
\providecommand \bibfield  [0]{\@secondoftwo}%
\providecommand \translation [1]{[#1]}%
\providecommand \BibitemOpen [0]{}%
\providecommand \bibitemStop [0]{}%
\providecommand \bibitemNoStop [0]{.\EOS\space}%
\providecommand \EOS [0]{\spacefactor3000\relax}%
\providecommand \BibitemShut  [1]{\csname bibitem#1\endcsname}%
\let\auto@bib@innerbib\@empty
\bibitem [{\citenamefont {Bonn}\ \emph {et~al.}(2017)\citenamefont {Bonn},
  \citenamefont {Denn}, \citenamefont {Berthier}, \citenamefont {Divoux},\ and\
  \citenamefont {Manneville}}]{Bonn:2017}%
  \BibitemOpen
  \bibfield  {author} {\bibinfo {author} {\bibfnamefont {D.}~\bibnamefont
  {Bonn}}, \bibinfo {author} {\bibfnamefont {M.~M.}\ \bibnamefont {Denn}},
  \bibinfo {author} {\bibfnamefont {L.}~\bibnamefont {Berthier}}, \bibinfo
  {author} {\bibfnamefont {T.}~\bibnamefont {Divoux}},\ and\ \bibinfo {author}
  {\bibfnamefont {S.}~\bibnamefont {Manneville}},\ }\bibfield  {title}
  {\bibinfo {title} {Yield stress materials in soft condensed matter},\
  }\href@noop {} {\bibfield  {journal} {\bibinfo  {journal} {Rev. Mod. Phys.}\
  }\textbf {\bibinfo {volume} {89}},\ \bibinfo {pages} {035005} (\bibinfo
  {year} {2017})}\BibitemShut {NoStop}%
\bibitem [{\citenamefont {Royall}\ \emph {et~al.}(2021)\citenamefont {Royall},
  \citenamefont {Faers}, \citenamefont {Fussell},\ and\ \citenamefont
  {Hallett}}]{Royall:2021}%
  \BibitemOpen
  \bibfield  {author} {\bibinfo {author} {\bibfnamefont {C.~P.}\ \bibnamefont
  {Royall}}, \bibinfo {author} {\bibfnamefont {M.~A.}\ \bibnamefont {Faers}},
  \bibinfo {author} {\bibfnamefont {S.~L.}\ \bibnamefont {Fussell}},\ and\
  \bibinfo {author} {\bibfnamefont {J.~E.}\ \bibnamefont {Hallett}},\
  }\bibfield  {title} {\bibinfo {title} {Real space analysis of colloidal gels:
  Triumphs, challenges and future directions},\ }\href@noop {} {\bibfield
  {journal} {\bibinfo  {journal} {J. Phys. Condens. Matter}\ }\textbf {\bibinfo
  {volume} {33}},\ \bibinfo {pages} {453002} (\bibinfo {year}
  {2021})}\BibitemShut {NoStop}%
\bibitem [{\citenamefont {Rueb}\ and\ \citenamefont
  {Zukoski}(1998)}]{rueb:1998}%
  \BibitemOpen
  \bibfield  {author} {\bibinfo {author} {\bibfnamefont {C.}~\bibnamefont
  {Rueb}}\ and\ \bibinfo {author} {\bibfnamefont {C.}~\bibnamefont {Zukoski}},\
  }\bibfield  {title} {\bibinfo {title} {Rheology of suspensions of weakly
  attractive particles: Approach to gelation},\ }\href@noop {} {\bibfield
  {journal} {\bibinfo  {journal} {J. Rheol.}\ }\textbf {\bibinfo {volume}
  {42}},\ \bibinfo {pages} {1451} (\bibinfo {year} {1998})}\BibitemShut
  {NoStop}%
\bibitem [{\citenamefont {Bergenholtz}\ \emph {et~al.}(2003)\citenamefont
  {Bergenholtz}, \citenamefont {Poon},\ and\ \citenamefont
  {Fuchs}}]{Bergenholtz:2003}%
  \BibitemOpen
  \bibfield  {author} {\bibinfo {author} {\bibfnamefont {J.}~\bibnamefont
  {Bergenholtz}}, \bibinfo {author} {\bibfnamefont {W.~C.}\ \bibnamefont
  {Poon}},\ and\ \bibinfo {author} {\bibfnamefont {M.}~\bibnamefont {Fuchs}},\
  }\bibfield  {title} {\bibinfo {title} {Gelation in model colloid-polymer
  mixtures},\ }\href@noop {} {\bibfield  {journal} {\bibinfo  {journal}
  {Langmuir}\ }\textbf {\bibinfo {volume} {19}},\ \bibinfo {pages} {4493}
  (\bibinfo {year} {2003})}\BibitemShut {NoStop}%
\bibitem [{\citenamefont {Varga}\ \emph {et~al.}(2015)\citenamefont {Varga},
  \citenamefont {Wang},\ and\ \citenamefont {Swan}}]{Varga:2015}%
  \BibitemOpen
  \bibfield  {author} {\bibinfo {author} {\bibfnamefont {Z.}~\bibnamefont
  {Varga}}, \bibinfo {author} {\bibfnamefont {G.}~\bibnamefont {Wang}},\ and\
  \bibinfo {author} {\bibfnamefont {J.}~\bibnamefont {Swan}},\ }\bibfield
  {title} {\bibinfo {title} {The hydrodynamics of colloidal gelation},\
  }\href@noop {} {\bibfield  {journal} {\bibinfo  {journal} {Soft Matter}\
  }\textbf {\bibinfo {volume} {11}},\ \bibinfo {pages} {9009} (\bibinfo {year}
  {2015})}\BibitemShut {NoStop}%
\bibitem [{\citenamefont {Rouwhorst}\ \emph {et~al.}(2020)\citenamefont
  {Rouwhorst}, \citenamefont {Schall}, \citenamefont {Ness}, \citenamefont
  {Blijdenstein},\ and\ \citenamefont {Zaccone}}]{Rouwhorst:2020}%
  \BibitemOpen
  \bibfield  {author} {\bibinfo {author} {\bibfnamefont {J.}~\bibnamefont
  {Rouwhorst}}, \bibinfo {author} {\bibfnamefont {P.}~\bibnamefont {Schall}},
  \bibinfo {author} {\bibfnamefont {C.}~\bibnamefont {Ness}}, \bibinfo {author}
  {\bibfnamefont {T.}~\bibnamefont {Blijdenstein}},\ and\ \bibinfo {author}
  {\bibfnamefont {A.}~\bibnamefont {Zaccone}},\ }\bibfield  {title} {\bibinfo
  {title} {Nonequilibrium master kinetic equation modeling of colloidal
  gelation},\ }\href@noop {} {\bibfield  {journal} {\bibinfo  {journal} {Phys.
  Rev. E}\ }\textbf {\bibinfo {volume} {102}},\ \bibinfo {pages} {022602}
  (\bibinfo {year} {2020})}\BibitemShut {NoStop}%
\bibitem [{\citenamefont {Morlet-Decarnin}\ \emph {et~al.}(2023)\citenamefont
  {Morlet-Decarnin}, \citenamefont {Divoux},\ and\ \citenamefont
  {Manneville}}]{MorletDecarnin:2023}%
  \BibitemOpen
  \bibfield  {author} {\bibinfo {author} {\bibfnamefont {L.}~\bibnamefont
  {Morlet-Decarnin}}, \bibinfo {author} {\bibfnamefont {T.}~\bibnamefont
  {Divoux}},\ and\ \bibinfo {author} {\bibfnamefont {S.}~\bibnamefont
  {Manneville}},\ }\bibfield  {title} {\bibinfo {title} {Critical-like gelation
  dynamics in cellulose nanocrystal suspensions},\ }\href@noop {} {\bibfield
  {journal} {\bibinfo  {journal} {ACS Macro Lett.}\ }\textbf {\bibinfo {volume}
  {12}},\ \bibinfo {pages} {1733} (\bibinfo {year} {2023})}\BibitemShut
  {NoStop}%
\bibitem [{\citenamefont {Aime}\ \emph {et~al.}(2018)\citenamefont {Aime},
  \citenamefont {Cipelletti},\ and\ \citenamefont {Ramos}}]{Aime:2018}%
  \BibitemOpen
  \bibfield  {author} {\bibinfo {author} {\bibfnamefont {S.}~\bibnamefont
  {Aime}}, \bibinfo {author} {\bibfnamefont {L.}~\bibnamefont {Cipelletti}},\
  and\ \bibinfo {author} {\bibfnamefont {L.}~\bibnamefont {Ramos}},\ }\bibfield
   {title} {\bibinfo {title} {Power law viscoelasticity of a fractal colloidal
  gel},\ }\href@noop {} {\bibfield  {journal} {\bibinfo  {journal} {J. Rheol.}\
  }\textbf {\bibinfo {volume} {62}},\ \bibinfo {pages} {1429} (\bibinfo {year}
  {2018})}\BibitemShut {NoStop}%
\bibitem [{\citenamefont {Keshavarz}\ \emph {et~al.}(2021)\citenamefont
  {Keshavarz}, \citenamefont {Rodrigues}, \citenamefont {Champenois},
  \citenamefont {Frith}, \citenamefont {Ilavsky}, \citenamefont {Geri},
  \citenamefont {Divoux}, \citenamefont {McKinley},\ and\ \citenamefont
  {Poulesquen}}]{Keshavarz:2021}%
  \BibitemOpen
  \bibfield  {author} {\bibinfo {author} {\bibfnamefont {B.}~\bibnamefont
  {Keshavarz}}, \bibinfo {author} {\bibfnamefont {D.~G.}\ \bibnamefont
  {Rodrigues}}, \bibinfo {author} {\bibfnamefont {J.-B.}\ \bibnamefont
  {Champenois}}, \bibinfo {author} {\bibfnamefont {M.~G.}\ \bibnamefont
  {Frith}}, \bibinfo {author} {\bibfnamefont {J.}~\bibnamefont {Ilavsky}},
  \bibinfo {author} {\bibfnamefont {M.}~\bibnamefont {Geri}}, \bibinfo {author}
  {\bibfnamefont {T.}~\bibnamefont {Divoux}}, \bibinfo {author} {\bibfnamefont
  {G.~H.}\ \bibnamefont {McKinley}},\ and\ \bibinfo {author} {\bibfnamefont
  {A.}~\bibnamefont {Poulesquen}},\ }\bibfield  {title} {\bibinfo {title}
  {Time--connectivity superposition and the gel/glass duality of weak colloidal
  gels},\ }\href@noop {} {\bibfield  {journal} {\bibinfo  {journal} {Proc.
  Natl. Acad. Sci. U.S.A.}\ }\textbf {\bibinfo {volume} {118}},\ \bibinfo
  {pages} {e2022339118} (\bibinfo {year} {2021})}\BibitemShut {NoStop}%
\bibitem [{\citenamefont {Johnson}\ \emph {et~al.}(2019)\citenamefont
  {Johnson}, \citenamefont {Zia}, \citenamefont {Moghimi},\ and\ \citenamefont
  {Petekidis}}]{Johnson:2019}%
  \BibitemOpen
  \bibfield  {author} {\bibinfo {author} {\bibfnamefont {L.~C.}\ \bibnamefont
  {Johnson}}, \bibinfo {author} {\bibfnamefont {R.~N.}\ \bibnamefont {Zia}},
  \bibinfo {author} {\bibfnamefont {E.}~\bibnamefont {Moghimi}},\ and\ \bibinfo
  {author} {\bibfnamefont {G.}~\bibnamefont {Petekidis}},\ }\bibfield  {title}
  {\bibinfo {title} {Influence of structure on the linear response rheology of
  colloidal gels},\ }\href@noop {} {\bibfield  {journal} {\bibinfo  {journal}
  {J. Rheol.}\ }\textbf {\bibinfo {volume} {63}},\ \bibinfo {pages} {583}
  (\bibinfo {year} {2019})}\BibitemShut {NoStop}%
\bibitem [{\citenamefont {Whitaker}\ \emph {et~al.}(2019)\citenamefont
  {Whitaker}, \citenamefont {Varga}, \citenamefont {Hsiao}, \citenamefont
  {Solomon}, \citenamefont {Swan},\ and\ \citenamefont
  {Furst}}]{Whitaker:2019}%
  \BibitemOpen
  \bibfield  {author} {\bibinfo {author} {\bibfnamefont {K.~A.}\ \bibnamefont
  {Whitaker}}, \bibinfo {author} {\bibfnamefont {Z.}~\bibnamefont {Varga}},
  \bibinfo {author} {\bibfnamefont {L.~C.}\ \bibnamefont {Hsiao}}, \bibinfo
  {author} {\bibfnamefont {M.~J.}\ \bibnamefont {Solomon}}, \bibinfo {author}
  {\bibfnamefont {J.~W.}\ \bibnamefont {Swan}},\ and\ \bibinfo {author}
  {\bibfnamefont {E.~M.}\ \bibnamefont {Furst}},\ }\bibfield  {title} {\bibinfo
  {title} {Colloidal gel elasticity arises from the packing of locally glassy
  clusters},\ }\href@noop {} {\bibfield  {journal} {\bibinfo  {journal} {Nat.
  Commun.}\ }\textbf {\bibinfo {volume} {10}},\ \bibinfo {pages} {2237}
  (\bibinfo {year} {2019})}\BibitemShut {NoStop}%
\bibitem [{\citenamefont {Kao}\ \emph {et~al.}(2022)\citenamefont {Kao},
  \citenamefont {Solomon},\ and\ \citenamefont {Ganesan}}]{Kao:2022}%
  \BibitemOpen
  \bibfield  {author} {\bibinfo {author} {\bibfnamefont {P.-K.}\ \bibnamefont
  {Kao}}, \bibinfo {author} {\bibfnamefont {M.~J.}\ \bibnamefont {Solomon}},\
  and\ \bibinfo {author} {\bibfnamefont {M.}~\bibnamefont {Ganesan}},\
  }\bibfield  {title} {\bibinfo {title} {Microstructure and elasticity of
  dilute gels of colloidal discoids},\ }\href@noop {} {\bibfield  {journal}
  {\bibinfo  {journal} {Soft Matter}\ }\textbf {\bibinfo {volume} {18}},\
  \bibinfo {pages} {1350} (\bibinfo {year} {2022})}\BibitemShut {NoStop}%
\bibitem [{\citenamefont {Bantawa}\ \emph {et~al.}(2023)\citenamefont
  {Bantawa}, \citenamefont {Keshavarz}, \citenamefont {Geri}, \citenamefont
  {Bouzid}, \citenamefont {Divoux}, \citenamefont {McKinley},\ and\
  \citenamefont {Del~Gado}}]{Bantawa:2023}%
  \BibitemOpen
  \bibfield  {author} {\bibinfo {author} {\bibfnamefont {M.}~\bibnamefont
  {Bantawa}}, \bibinfo {author} {\bibfnamefont {B.}~\bibnamefont {Keshavarz}},
  \bibinfo {author} {\bibfnamefont {M.}~\bibnamefont {Geri}}, \bibinfo {author}
  {\bibfnamefont {M.}~\bibnamefont {Bouzid}}, \bibinfo {author} {\bibfnamefont
  {T.}~\bibnamefont {Divoux}}, \bibinfo {author} {\bibfnamefont {G.~H.}\
  \bibnamefont {McKinley}},\ and\ \bibinfo {author} {\bibfnamefont
  {E.}~\bibnamefont {Del~Gado}},\ }\bibfield  {title} {\bibinfo {title} {The
  hidden hierarchical nature of soft particulate gels},\ }\href@noop {}
  {\bibfield  {journal} {\bibinfo  {journal} {Nature Phys.}\ }\textbf {\bibinfo
  {volume} {19}},\ \bibinfo {pages} {1178} (\bibinfo {year}
  {2023})}\BibitemShut {NoStop}%
\bibitem [{\citenamefont {Richard}\ and\ \citenamefont
  {Bouzid}(2025)}]{Richard:2025}%
  \BibitemOpen
  \bibfield  {author} {\bibinfo {author} {\bibfnamefont {D.}~\bibnamefont
  {Richard}}\ and\ \bibinfo {author} {\bibfnamefont {M.}~\bibnamefont
  {Bouzid}},\ }\bibfield  {title} {\bibinfo {title} {How rigidity percolation
  and bending stiffness shape colloidal gel elasticity},\ }\href@noop {}
  {\bibfield  {journal} {\bibinfo  {journal} {arXiv preprint arXiv:2504.19568}\
  } (\bibinfo {year} {2025})}\BibitemShut {NoStop}%
\bibitem [{\citenamefont {Fournier}\ \emph {et~al.}(2024)\citenamefont
  {Fournier}, \citenamefont {Chevalier}, \citenamefont {Perez-Robles},
  \citenamefont {Carotenuto}, \citenamefont {Minale}, \citenamefont {Reveron},\
  and\ \citenamefont {Baeza}}]{Fournier:2024}%
  \BibitemOpen
  \bibfield  {author} {\bibinfo {author} {\bibfnamefont {S.}~\bibnamefont
  {Fournier}}, \bibinfo {author} {\bibfnamefont {J.}~\bibnamefont {Chevalier}},
  \bibinfo {author} {\bibfnamefont {S.}~\bibnamefont {Perez-Robles}}, \bibinfo
  {author} {\bibfnamefont {C.}~\bibnamefont {Carotenuto}}, \bibinfo {author}
  {\bibfnamefont {M.}~\bibnamefont {Minale}}, \bibinfo {author} {\bibfnamefont
  {H.}~\bibnamefont {Reveron}},\ and\ \bibinfo {author} {\bibfnamefont {G.~P.}\
  \bibnamefont {Baeza}},\ }\bibfield  {title} {\bibinfo {title} {Spreading
  ceramic stereolithography pastes: Insights from shear-and
  orthogonal-rheology},\ }\href@noop {} {\bibfield  {journal} {\bibinfo
  {journal} {J. Rheol.}\ }\textbf {\bibinfo {volume} {68}},\ \bibinfo {pages}
  {83} (\bibinfo {year} {2024})}\BibitemShut {NoStop}%
\bibitem [{\citenamefont {Hanley}\ \emph {et~al.}(1999)\citenamefont {Hanley},
  \citenamefont {Muzny}, \citenamefont {Butler}, \citenamefont {Straty},
  \citenamefont {Bartlett},\ and\ \citenamefont {Drabarek}}]{Hanley:1999}%
  \BibitemOpen
  \bibfield  {author} {\bibinfo {author} {\bibfnamefont {H.}~\bibnamefont
  {Hanley}}, \bibinfo {author} {\bibfnamefont {C.}~\bibnamefont {Muzny}},
  \bibinfo {author} {\bibfnamefont {B.}~\bibnamefont {Butler}}, \bibinfo
  {author} {\bibfnamefont {G.}~\bibnamefont {Straty}}, \bibinfo {author}
  {\bibfnamefont {J.}~\bibnamefont {Bartlett}},\ and\ \bibinfo {author}
  {\bibfnamefont {E.}~\bibnamefont {Drabarek}},\ }\bibfield  {title} {\bibinfo
  {title} {Shear-induced restructuring of concentrated colloidal silica gels},\
  }\href@noop {} {\bibfield  {journal} {\bibinfo  {journal} {J. Condens. Matter
  Phys.}\ }\textbf {\bibinfo {volume} {11}},\ \bibinfo {pages} {1369} (\bibinfo
  {year} {1999})}\BibitemShut {NoStop}%
\bibitem [{\citenamefont {Mokhtari}\ \emph {et~al.}(2008)\citenamefont
  {Mokhtari}, \citenamefont {Chakrabarti}, \citenamefont {Sorensen},
  \citenamefont {Cheng},\ and\ \citenamefont {Vigil}}]{Mokhtari:2008}%
  \BibitemOpen
  \bibfield  {author} {\bibinfo {author} {\bibfnamefont {T.}~\bibnamefont
  {Mokhtari}}, \bibinfo {author} {\bibfnamefont {A.}~\bibnamefont
  {Chakrabarti}}, \bibinfo {author} {\bibfnamefont {C.~M.}\ \bibnamefont
  {Sorensen}}, \bibinfo {author} {\bibfnamefont {C.-Y.}\ \bibnamefont
  {Cheng}},\ and\ \bibinfo {author} {\bibfnamefont {D.}~\bibnamefont {Vigil}},\
  }\bibfield  {title} {\bibinfo {title} {The effect of shear on colloidal
  aggregation and gelation studied using small-angle light scattering},\
  }\href@noop {} {\bibfield  {journal} {\bibinfo  {journal} {J. Colloid
  Interface Sci.}\ }\textbf {\bibinfo {volume} {327}},\ \bibinfo {pages} {216 }
  (\bibinfo {year} {2008})}\BibitemShut {NoStop}%
\bibitem [{\citenamefont {Koumakis}\ \emph {et~al.}(2015)\citenamefont
  {Koumakis}, \citenamefont {Moghimi}, \citenamefont {Besseling}, \citenamefont
  {Poon}, \citenamefont {Brady},\ and\ \citenamefont
  {Petekidis}}]{Koumakis:2015}%
  \BibitemOpen
  \bibfield  {author} {\bibinfo {author} {\bibfnamefont {N.}~\bibnamefont
  {Koumakis}}, \bibinfo {author} {\bibfnamefont {E.}~\bibnamefont {Moghimi}},
  \bibinfo {author} {\bibfnamefont {R.}~\bibnamefont {Besseling}}, \bibinfo
  {author} {\bibfnamefont {W.~C.}\ \bibnamefont {Poon}}, \bibinfo {author}
  {\bibfnamefont {J.~F.}\ \bibnamefont {Brady}},\ and\ \bibinfo {author}
  {\bibfnamefont {G.}~\bibnamefont {Petekidis}},\ }\bibfield  {title} {\bibinfo
  {title} {Tuning colloidal gels by shear},\ }\href@noop {} {\bibfield
  {journal} {\bibinfo  {journal} {Soft Matter}\ }\textbf {\bibinfo {volume}
  {11}},\ \bibinfo {pages} {4640} (\bibinfo {year} {2015})}\BibitemShut
  {NoStop}%
\bibitem [{\citenamefont {Moghimi}\ \emph {et~al.}(2017)\citenamefont
  {Moghimi}, \citenamefont {Jacob}, \citenamefont {Koumakis},\ and\
  \citenamefont {Petekidis}}]{Moghimi:2017}%
  \BibitemOpen
  \bibfield  {author} {\bibinfo {author} {\bibfnamefont {E.}~\bibnamefont
  {Moghimi}}, \bibinfo {author} {\bibfnamefont {A.~R.}\ \bibnamefont {Jacob}},
  \bibinfo {author} {\bibfnamefont {N.}~\bibnamefont {Koumakis}},\ and\
  \bibinfo {author} {\bibfnamefont {G.}~\bibnamefont {Petekidis}},\ }\bibfield
  {title} {\bibinfo {title} {Colloidal gels tuned by oscillatory shear},\
  }\href@noop {} {\bibfield  {journal} {\bibinfo  {journal} {Soft Matter}\
  }\textbf {\bibinfo {volume} {13}},\ \bibinfo {pages} {2371} (\bibinfo {year}
  {2017})}\BibitemShut {NoStop}%
\bibitem [{\citenamefont {Sudreau}\ \emph
  {et~al.}(2022{\natexlab{a}})\citenamefont {Sudreau}, \citenamefont
  {Manneville}, \citenamefont {Servel},\ and\ \citenamefont
  {Divoux}}]{Sudreau:2022}%
  \BibitemOpen
  \bibfield  {author} {\bibinfo {author} {\bibfnamefont {I.}~\bibnamefont
  {Sudreau}}, \bibinfo {author} {\bibfnamefont {S.}~\bibnamefont {Manneville}},
  \bibinfo {author} {\bibfnamefont {M.}~\bibnamefont {Servel}},\ and\ \bibinfo
  {author} {\bibfnamefont {T.}~\bibnamefont {Divoux}},\ }\bibfield  {title}
  {\bibinfo {title} {Shear-induced memory effects in boehmite gels},\
  }\href@noop {} {\bibfield  {journal} {\bibinfo  {journal} {J. Rheol.}\
  }\textbf {\bibinfo {volume} {66}},\ \bibinfo {pages} {91} (\bibinfo {year}
  {2022}{\natexlab{a}})}\BibitemShut {NoStop}%
\bibitem [{\citenamefont {Muzny}\ \emph {et~al.}(2023)\citenamefont {Muzny},
  \citenamefont {de~Campo}, \citenamefont {Sokolova}, \citenamefont {Garvey},
  \citenamefont {Rehm},\ and\ \citenamefont {Hanley}}]{Muzny:2023}%
  \BibitemOpen
  \bibfield  {author} {\bibinfo {author} {\bibfnamefont {C.}~\bibnamefont
  {Muzny}}, \bibinfo {author} {\bibfnamefont {L.}~\bibnamefont {de~Campo}},
  \bibinfo {author} {\bibfnamefont {A.}~\bibnamefont {Sokolova}}, \bibinfo
  {author} {\bibfnamefont {C.~J.}\ \bibnamefont {Garvey}}, \bibinfo {author}
  {\bibfnamefont {C.}~\bibnamefont {Rehm}},\ and\ \bibinfo {author}
  {\bibfnamefont {H.}~\bibnamefont {Hanley}},\ }\bibfield  {title} {\bibinfo
  {title} {Shear influence on colloidal cluster growth: a {SANS} and {USANS}
  study},\ }\href@noop {} {\bibfield  {journal} {\bibinfo  {journal} {J. Appl.
  Crystallogr.}\ }\textbf {\bibinfo {volume} {56}},\ \bibinfo {pages} {1371}
  (\bibinfo {year} {2023})}\BibitemShut {NoStop}%
\bibitem [{\citenamefont {Seljelid}\ \emph {et~al.}(2024)\citenamefont
  {Seljelid}, \citenamefont {Neto}, \citenamefont {Akanno}, \citenamefont
  {Ceccato}, \citenamefont {Ravindranathan}, \citenamefont {Azmi},
  \citenamefont {Cavalcanti}, \citenamefont {Fjelde}, \citenamefont {Knudsen},\
  and\ \citenamefont {Fossum}}]{Seljelid:2024}%
  \BibitemOpen
  \bibfield  {author} {\bibinfo {author} {\bibfnamefont {K.~K.}\ \bibnamefont
  {Seljelid}}, \bibinfo {author} {\bibfnamefont {O.~T.}\ \bibnamefont {Neto}},
  \bibinfo {author} {\bibfnamefont {A.~N.}\ \bibnamefont {Akanno}}, \bibinfo
  {author} {\bibfnamefont {B.~T.}\ \bibnamefont {Ceccato}}, \bibinfo {author}
  {\bibfnamefont {R.~P.}\ \bibnamefont {Ravindranathan}}, \bibinfo {author}
  {\bibfnamefont {N.}~\bibnamefont {Azmi}}, \bibinfo {author} {\bibfnamefont
  {L.~P.}\ \bibnamefont {Cavalcanti}}, \bibinfo {author} {\bibfnamefont
  {I.}~\bibnamefont {Fjelde}}, \bibinfo {author} {\bibfnamefont {K.~D.}\
  \bibnamefont {Knudsen}},\ and\ \bibinfo {author} {\bibfnamefont {J.~O.}\
  \bibnamefont {Fossum}},\ }\bibfield  {title} {\bibinfo {title} {Growth
  kinetics and structure of a colloidal silica-based network: in situ
  {R}heo{SAXS} investigations},\ }\href@noop {} {\bibfield  {journal} {\bibinfo
   {journal} {Eur. Phys. J.: Spec. Top.}\ }\textbf {\bibinfo {volume} {233}},\
  \bibinfo {pages} {2757} (\bibinfo {year} {2024})}\BibitemShut {NoStop}%
\bibitem [{\citenamefont {Bhaumik}\ \emph
  {et~al.}(2025{\natexlab{a}})\citenamefont {Bhaumik}, \citenamefont
  {Liverpool}, \citenamefont {Royall},\ and\ \citenamefont
  {Jack}}]{Bhaumik:2025a}%
  \BibitemOpen
  \bibfield  {author} {\bibinfo {author} {\bibfnamefont {H.}~\bibnamefont
  {Bhaumik}}, \bibinfo {author} {\bibfnamefont {T.~B.}\ \bibnamefont
  {Liverpool}}, \bibinfo {author} {\bibfnamefont {C.~P.}\ \bibnamefont
  {Royall}},\ and\ \bibinfo {author} {\bibfnamefont {R.~L.}\ \bibnamefont
  {Jack}},\ }\bibfield  {title} {\bibinfo {title} {Yielding in colloidal gels:
  From local structure to mesoscale strand breakage and macroscopic failure},\
  }\href@noop {} {\bibfield  {journal} {\bibinfo  {journal} {Phys. Rev. E}\
  }\textbf {\bibinfo {volume} {111}},\ \bibinfo {pages} {055412} (\bibinfo
  {year} {2025}{\natexlab{a}})}\BibitemShut {NoStop}%
\bibitem [{\citenamefont {Bhaumik}\ \emph
  {et~al.}(2025{\natexlab{b}})\citenamefont {Bhaumik}, \citenamefont {Hallett},
  \citenamefont {Liverpool}, \citenamefont {Jack},\ and\ \citenamefont
  {Royall}}]{Bhaumik:2025b}%
  \BibitemOpen
  \bibfield  {author} {\bibinfo {author} {\bibfnamefont {H.}~\bibnamefont
  {Bhaumik}}, \bibinfo {author} {\bibfnamefont {J.}~\bibnamefont {Hallett}},
  \bibinfo {author} {\bibfnamefont {T.}~\bibnamefont {Liverpool}}, \bibinfo
  {author} {\bibfnamefont {R.}~\bibnamefont {Jack}},\ and\ \bibinfo {author}
  {\bibfnamefont {C.}~\bibnamefont {Royall}},\ }\bibfield  {title} {\bibinfo
  {title} {Cyclically sheared colloidal gels: structural change and delayed
  failure time},\ }\href {https://doi.org/10.1039/d5sm00647c} {\bibfield
  {journal} {\bibinfo  {journal} {Soft Matter}\ }\textbf {\bibinfo {volume}
  {21}},\ \bibinfo {pages} {8555} (\bibinfo {year}
  {2025}{\natexlab{b}})}\BibitemShut {NoStop}%
\bibitem [{\citenamefont {Gibaud}\ \emph {et~al.}(2020)\citenamefont {Gibaud},
  \citenamefont {Dag\`es}, \citenamefont {Lidon}, \citenamefont {Jung},
  \citenamefont {Ahour\'e}, \citenamefont {Sztucki}, \citenamefont
  {Poulesquen}, \citenamefont {Hengl}, \citenamefont {Pignon},\ and\
  \citenamefont {Manneville}}]{Gibaud:2020}%
  \BibitemOpen
  \bibfield  {author} {\bibinfo {author} {\bibfnamefont {T.}~\bibnamefont
  {Gibaud}}, \bibinfo {author} {\bibfnamefont {N.}~\bibnamefont {Dag\`es}},
  \bibinfo {author} {\bibfnamefont {P.}~\bibnamefont {Lidon}}, \bibinfo
  {author} {\bibfnamefont {G.}~\bibnamefont {Jung}}, \bibinfo {author}
  {\bibfnamefont {L.~C.}\ \bibnamefont {Ahour\'e}}, \bibinfo {author}
  {\bibfnamefont {M.}~\bibnamefont {Sztucki}}, \bibinfo {author} {\bibfnamefont
  {A.}~\bibnamefont {Poulesquen}}, \bibinfo {author} {\bibfnamefont
  {N.}~\bibnamefont {Hengl}}, \bibinfo {author} {\bibfnamefont
  {F.}~\bibnamefont {Pignon}},\ and\ \bibinfo {author} {\bibfnamefont
  {S.}~\bibnamefont {Manneville}},\ }\bibfield  {title} {\bibinfo {title}
  {Rheoacoustic gels: Tuning mechanical and flow properties of colloidal gels
  with ultrasonic vibrations},\ }\href@noop {} {\bibfield  {journal} {\bibinfo
  {journal} {Phys. Rev. X}\ }\textbf {\bibinfo {volume} {10}},\ \bibinfo
  {pages} {011028} (\bibinfo {year} {2020})}\BibitemShut {NoStop}%
\bibitem [{\citenamefont {Dag{\`e}s}\ \emph {et~al.}(2021)\citenamefont
  {Dag{\`e}s}, \citenamefont {Lidon}, \citenamefont {Jung}, \citenamefont
  {Pignon}, \citenamefont {Manneville},\ and\ \citenamefont
  {Gibaud}}]{Dages:2021}%
  \BibitemOpen
  \bibfield  {author} {\bibinfo {author} {\bibfnamefont {N.}~\bibnamefont
  {Dag{\`e}s}}, \bibinfo {author} {\bibfnamefont {P.}~\bibnamefont {Lidon}},
  \bibinfo {author} {\bibfnamefont {G.}~\bibnamefont {Jung}}, \bibinfo {author}
  {\bibfnamefont {F.}~\bibnamefont {Pignon}}, \bibinfo {author} {\bibfnamefont
  {S.}~\bibnamefont {Manneville}},\ and\ \bibinfo {author} {\bibfnamefont
  {T.}~\bibnamefont {Gibaud}},\ }\bibfield  {title} {\bibinfo {title}
  {Mechanics and structure of carbon black gels under high-power ultrasound},\
  }\href@noop {} {\bibfield  {journal} {\bibinfo  {journal} {J. Rheol.}\
  }\textbf {\bibinfo {volume} {65}},\ \bibinfo {pages} {477} (\bibinfo {year}
  {2021})}\BibitemShut {NoStop}%
\bibitem [{\citenamefont {Saint-Michel}\ \emph {et~al.}(2022)\citenamefont
  {Saint-Michel}, \citenamefont {Petekidis},\ and\ \citenamefont
  {Garbin}}]{SaintMichel:2022}%
  \BibitemOpen
  \bibfield  {author} {\bibinfo {author} {\bibfnamefont {B.}~\bibnamefont
  {Saint-Michel}}, \bibinfo {author} {\bibfnamefont {G.}~\bibnamefont
  {Petekidis}},\ and\ \bibinfo {author} {\bibfnamefont {V.}~\bibnamefont
  {Garbin}},\ }\bibfield  {title} {\bibinfo {title} {Tuning local
  microstructure of colloidal gels by ultrasound-activated deformable
  inclusions},\ }\href@noop {} {\bibfield  {journal} {\bibinfo  {journal} {Soft
  Matter}\ }\textbf {\bibinfo {volume} {18}},\ \bibinfo {pages} {2092}
  (\bibinfo {year} {2022})}\BibitemShut {NoStop}%
\bibitem [{\citenamefont {Gadige}\ and\ \citenamefont
  {Bandyopadhyay}(2018)}]{Gadige:2018}%
  \BibitemOpen
  \bibfield  {author} {\bibinfo {author} {\bibfnamefont {P.}~\bibnamefont
  {Gadige}}\ and\ \bibinfo {author} {\bibfnamefont {R.}~\bibnamefont
  {Bandyopadhyay}},\ }\bibfield  {title} {\bibinfo {title} {Electric field
  induced gelation in aqueous nanoclay suspensions},\ }\href@noop {} {\bibfield
   {journal} {\bibinfo  {journal} {Soft Matter}\ }\textbf {\bibinfo {volume}
  {14}},\ \bibinfo {pages} {6974} (\bibinfo {year} {2018})}\BibitemShut
  {NoStop}%
\bibitem [{\citenamefont {Semwal}\ \emph {et~al.}(2022)\citenamefont {Semwal},
  \citenamefont {Clowe-Coish}, \citenamefont {Saika-Voivod},\ and\
  \citenamefont {Yethiraj}}]{Semwal:2022}%
  \BibitemOpen
  \bibfield  {author} {\bibinfo {author} {\bibfnamefont {S.}~\bibnamefont
  {Semwal}}, \bibinfo {author} {\bibfnamefont {C.}~\bibnamefont {Clowe-Coish}},
  \bibinfo {author} {\bibfnamefont {I.}~\bibnamefont {Saika-Voivod}},\ and\
  \bibinfo {author} {\bibfnamefont {A.}~\bibnamefont {Yethiraj}},\ }\bibfield
  {title} {\bibinfo {title} {Tunable colloids with dipolar and depletion
  interactions: Toward field-switchable crystals and gels},\ }\href@noop {}
  {\bibfield  {journal} {\bibinfo  {journal} {Phys. Rev. X}\ }\textbf {\bibinfo
  {volume} {12}},\ \bibinfo {pages} {041021} (\bibinfo {year}
  {2022})}\BibitemShut {NoStop}%
\bibitem [{\citenamefont {Munteanu}\ \emph {et~al.}(2025)\citenamefont
  {Munteanu}, \citenamefont {Munteanu},\ and\ \citenamefont
  {Sedlacik}}]{Munteanu:2025}%
  \BibitemOpen
  \bibfield  {author} {\bibinfo {author} {\bibfnamefont {L.}~\bibnamefont
  {Munteanu}}, \bibinfo {author} {\bibfnamefont {A.}~\bibnamefont {Munteanu}},\
  and\ \bibinfo {author} {\bibfnamefont {M.}~\bibnamefont {Sedlacik}},\
  }\bibfield  {title} {\bibinfo {title} {Electrorheological fluids: A living
  review},\ }\href@noop {} {\bibfield  {journal} {\bibinfo  {journal} {Progress
  in Materials Science}\ }\textbf {\bibinfo {volume} {151}},\ \bibinfo {pages}
  {101421} (\bibinfo {year} {2025})}\BibitemShut {NoStop}%
\bibitem [{\citenamefont {Selomulya}\ \emph {et~al.}(2001)\citenamefont
  {Selomulya}, \citenamefont {Amal}, \citenamefont {Bushell},\ and\
  \citenamefont {Waite}}]{Selomulya:2001}%
  \BibitemOpen
  \bibfield  {author} {\bibinfo {author} {\bibfnamefont {C.}~\bibnamefont
  {Selomulya}}, \bibinfo {author} {\bibfnamefont {R.}~\bibnamefont {Amal}},
  \bibinfo {author} {\bibfnamefont {G.}~\bibnamefont {Bushell}},\ and\ \bibinfo
  {author} {\bibfnamefont {T.}~\bibnamefont {Waite}},\ }\bibfield  {title}
  {\bibinfo {title} {Evidence of shear rate dependence on restructuring and
  breakup of latex aggregates},\ }\href@noop {} {\bibfield  {journal} {\bibinfo
   {journal} {J. Colloid Interface Sci.}\ }\textbf {\bibinfo {volume} {236}},\
  \bibinfo {pages} {67} (\bibinfo {year} {2001})}\BibitemShut {NoStop}%
\bibitem [{\citenamefont {Selomulya}\ \emph {et~al.}(2002)\citenamefont
  {Selomulya}, \citenamefont {Bushell}, \citenamefont {Amal},\ and\
  \citenamefont {Waite}}]{Selomulya:2002}%
  \BibitemOpen
  \bibfield  {author} {\bibinfo {author} {\bibfnamefont {C.}~\bibnamefont
  {Selomulya}}, \bibinfo {author} {\bibfnamefont {G.}~\bibnamefont {Bushell}},
  \bibinfo {author} {\bibfnamefont {R.}~\bibnamefont {Amal}},\ and\ \bibinfo
  {author} {\bibfnamefont {T.~D.}\ \bibnamefont {Waite}},\ }\bibfield  {title}
  {\bibinfo {title} {Aggregation mechanisms of latex of different particle
  sizes in a controlled shear environment},\ }\href@noop {} {\bibfield
  {journal} {\bibinfo  {journal} {Langmuir}\ }\textbf {\bibinfo {volume}
  {18}},\ \bibinfo {pages} {1974} (\bibinfo {year} {2002})}\BibitemShut
  {NoStop}%
\bibitem [{\citenamefont {Eggersdorfer}\ \emph {et~al.}(2010)\citenamefont
  {Eggersdorfer}, \citenamefont {Kadau}, \citenamefont {Herrmann},\ and\
  \citenamefont {Pratsinis}}]{Eggersdorfer:2010}%
  \BibitemOpen
  \bibfield  {author} {\bibinfo {author} {\bibfnamefont {M.}~\bibnamefont
  {Eggersdorfer}}, \bibinfo {author} {\bibfnamefont {D.}~\bibnamefont {Kadau}},
  \bibinfo {author} {\bibfnamefont {H.~J.}\ \bibnamefont {Herrmann}},\ and\
  \bibinfo {author} {\bibfnamefont {S.~E.}\ \bibnamefont {Pratsinis}},\
  }\bibfield  {title} {\bibinfo {title} {Fragmentation and restructuring of
  soft-agglomerates under shear},\ }\href@noop {} {\bibfield  {journal}
  {\bibinfo  {journal} {J. Colloid Interface Sci.}\ }\textbf {\bibinfo {volume}
  {342}},\ \bibinfo {pages} {261} (\bibinfo {year} {2010})}\BibitemShut
  {NoStop}%
\bibitem [{\citenamefont {Lieu}\ and\ \citenamefont
  {Harada}(2016)}]{Lieu:2016}%
  \BibitemOpen
  \bibfield  {author} {\bibinfo {author} {\bibfnamefont {U.~T.}\ \bibnamefont
  {Lieu}}\ and\ \bibinfo {author} {\bibfnamefont {S.}~\bibnamefont {Harada}},\
  }\bibfield  {title} {\bibinfo {title} {Restructuring capability of
  non-fractal aggregate in simple shear flow},\ }\href
  {https://doi.org/https://doi.org/10.1016/j.apt.2016.03.009} {\bibfield
  {journal} {\bibinfo  {journal} {Adv. Powder Technol.}\ }\textbf {\bibinfo
  {volume} {27}},\ \bibinfo {pages} {1037 } (\bibinfo {year}
  {2016})}\BibitemShut {NoStop}%
\bibitem [{\citenamefont {Bauland}\ \emph {et~al.}(2024)\citenamefont
  {Bauland}, \citenamefont {Bouthier}, \citenamefont {Poulesquen},\ and\
  \citenamefont {Gibaud}}]{Bauland2024}%
  \BibitemOpen
  \bibfield  {author} {\bibinfo {author} {\bibfnamefont {J.}~\bibnamefont
  {Bauland}}, \bibinfo {author} {\bibfnamefont {L.-V.}\ \bibnamefont
  {Bouthier}}, \bibinfo {author} {\bibfnamefont {A.}~\bibnamefont
  {Poulesquen}},\ and\ \bibinfo {author} {\bibfnamefont {T.}~\bibnamefont
  {Gibaud}},\ }\bibfield  {title} {\bibinfo {title} {Attractive carbon black
  dispersions: Structural and mechanical responses to shear},\ }\href@noop {}
  {\bibfield  {journal} {\bibinfo  {journal} {J. Rheol.}\ }\textbf {\bibinfo
  {volume} {68}},\ \bibinfo {pages} {429} (\bibinfo {year} {2024})}\BibitemShut
  {NoStop}%
\bibitem [{\citenamefont {Varga}\ and\ \citenamefont
  {Swan}(2018)}]{Varga:2018}%
  \BibitemOpen
  \bibfield  {author} {\bibinfo {author} {\bibfnamefont {Z.}~\bibnamefont
  {Varga}}\ and\ \bibinfo {author} {\bibfnamefont {J.~W.}\ \bibnamefont
  {Swan}},\ }\bibfield  {title} {\bibinfo {title} {Large scale anisotropies in
  sheared colloidal gels},\ }\href@noop {} {\bibfield  {journal} {\bibinfo
  {journal} {J. Rheol.}\ }\textbf {\bibinfo {volume} {62}},\ \bibinfo {pages}
  {405} (\bibinfo {year} {2018})}\BibitemShut {NoStop}%
\bibitem [{\citenamefont {Varga}\ \emph {et~al.}(2019)\citenamefont {Varga},
  \citenamefont {Grenard}, \citenamefont {Pecorario}, \citenamefont {Taberlet},
  \citenamefont {Dolique}, \citenamefont {Manneville}, \citenamefont {Divoux},
  \citenamefont {McKinley},\ and\ \citenamefont {Swan}}]{Varga:2019}%
  \BibitemOpen
  \bibfield  {author} {\bibinfo {author} {\bibfnamefont {Z.}~\bibnamefont
  {Varga}}, \bibinfo {author} {\bibfnamefont {V.}~\bibnamefont {Grenard}},
  \bibinfo {author} {\bibfnamefont {S.}~\bibnamefont {Pecorario}}, \bibinfo
  {author} {\bibfnamefont {N.}~\bibnamefont {Taberlet}}, \bibinfo {author}
  {\bibfnamefont {V.}~\bibnamefont {Dolique}}, \bibinfo {author} {\bibfnamefont
  {S.}~\bibnamefont {Manneville}}, \bibinfo {author} {\bibfnamefont
  {T.}~\bibnamefont {Divoux}}, \bibinfo {author} {\bibfnamefont {G.~H.}\
  \bibnamefont {McKinley}},\ and\ \bibinfo {author} {\bibfnamefont {J.~W.}\
  \bibnamefont {Swan}},\ }\bibfield  {title} {\bibinfo {title} {Hydrodynamics
  control shear-induced pattern formation in attractive suspensions},\
  }\href@noop {} {\bibfield  {journal} {\bibinfo  {journal} {Proc. Natl. Acad.
  Sci}\ }\textbf {\bibinfo {volume} {116}},\ \bibinfo {pages} {12193} (\bibinfo
  {year} {2019})}\BibitemShut {NoStop}%
\bibitem [{\citenamefont {Das}\ \emph {et~al.}(2021)\citenamefont {Das},
  \citenamefont {Chambon}, \citenamefont {Varga}, \citenamefont {Vamvakaki},
  \citenamefont {Swan},\ and\ \citenamefont {Petekidis}}]{Das:2021}%
  \BibitemOpen
  \bibfield  {author} {\bibinfo {author} {\bibfnamefont {M.}~\bibnamefont
  {Das}}, \bibinfo {author} {\bibfnamefont {L.}~\bibnamefont {Chambon}},
  \bibinfo {author} {\bibfnamefont {Z.}~\bibnamefont {Varga}}, \bibinfo
  {author} {\bibfnamefont {M.}~\bibnamefont {Vamvakaki}}, \bibinfo {author}
  {\bibfnamefont {J.~W.}\ \bibnamefont {Swan}},\ and\ \bibinfo {author}
  {\bibfnamefont {G.}~\bibnamefont {Petekidis}},\ }\bibfield  {title} {\bibinfo
  {title} {Shear driven vorticity aligned flocs in a suspension of attractive
  rigid rods},\ }\href@noop {} {\bibfield  {journal} {\bibinfo  {journal} {Soft
  Matter}\ }\textbf {\bibinfo {volume} {17}},\ \bibinfo {pages} {1232}
  (\bibinfo {year} {2021})}\BibitemShut {NoStop}%
\bibitem [{\citenamefont {Bauland}\ and\ \citenamefont
  {Gibaud}(2025)}]{Bauland2025}%
  \BibitemOpen
  \bibfield  {author} {\bibinfo {author} {\bibfnamefont {J.}~\bibnamefont
  {Bauland}}\ and\ \bibinfo {author} {\bibfnamefont {T.}~\bibnamefont
  {Gibaud}},\ }\bibfield  {title} {\bibinfo {title} {Shear-driven memory
  effects in carbon black gels},\ }\href@noop {} {\bibfield  {journal}
  {\bibinfo  {journal} {arXiv preprint arXiv:2508.02239}\ } (\bibinfo {year}
  {2025})}\BibitemShut {NoStop}%
\bibitem [{\citenamefont {Jamali}\ \emph {et~al.}(2020)\citenamefont {Jamali},
  \citenamefont {Armstrong},\ and\ \citenamefont {McKinley}}]{Jamali:2020}%
  \BibitemOpen
  \bibfield  {author} {\bibinfo {author} {\bibfnamefont {S.}~\bibnamefont
  {Jamali}}, \bibinfo {author} {\bibfnamefont {R.~C.}\ \bibnamefont
  {Armstrong}},\ and\ \bibinfo {author} {\bibfnamefont {G.~H.}\ \bibnamefont
  {McKinley}},\ }\bibfield  {title} {\bibinfo {title} {Time-rate-transformation
  framework for targeted assembly of short-range attractive colloidal
  suspensions},\ }\href@noop {} {\bibfield  {journal} {\bibinfo  {journal}
  {Mater. Today Adv.}\ }\textbf {\bibinfo {volume} {5}},\ \bibinfo {pages}
  {100026} (\bibinfo {year} {2020})}\BibitemShut {NoStop}%
\bibitem [{\citenamefont {Schwen}\ \emph {et~al.}(2020)\citenamefont {Schwen},
  \citenamefont {Ramaswamy}, \citenamefont {Cheng}, \citenamefont {Jan},\ and\
  \citenamefont {Cohen}}]{Schwen:2020}%
  \BibitemOpen
  \bibfield  {author} {\bibinfo {author} {\bibfnamefont {E.~M.}\ \bibnamefont
  {Schwen}}, \bibinfo {author} {\bibfnamefont {M.}~\bibnamefont {Ramaswamy}},
  \bibinfo {author} {\bibfnamefont {C.-M.}\ \bibnamefont {Cheng}}, \bibinfo
  {author} {\bibfnamefont {L.}~\bibnamefont {Jan}},\ and\ \bibinfo {author}
  {\bibfnamefont {I.}~\bibnamefont {Cohen}},\ }\bibfield  {title} {\bibinfo
  {title} {Embedding orthogonal memories in a colloidal gel through oscillatory
  shear},\ }\href@noop {} {\bibfield  {journal} {\bibinfo  {journal} {Soft
  Matter}\ }\textbf {\bibinfo {volume} {16}},\ \bibinfo {pages} {3746}
  (\bibinfo {year} {2020})}\BibitemShut {NoStop}%
\bibitem [{\citenamefont {Das}\ and\ \citenamefont
  {Petekidis}(2022)}]{Das:2022}%
  \BibitemOpen
  \bibfield  {author} {\bibinfo {author} {\bibfnamefont {M.}~\bibnamefont
  {Das}}\ and\ \bibinfo {author} {\bibfnamefont {G.}~\bibnamefont
  {Petekidis}},\ }\bibfield  {title} {\bibinfo {title} {Shear induced tuning
  and memory effects in colloidal gels of rods and spheres},\ }\href@noop {}
  {\bibfield  {journal} {\bibinfo  {journal} {J. Chem. Phys.}\ }\textbf
  {\bibinfo {volume} {157}} (\bibinfo {year} {2022})}\BibitemShut {NoStop}%
\bibitem [{\citenamefont {Sudreau}\ \emph {et~al.}(2023)\citenamefont
  {Sudreau}, \citenamefont {Servel}, \citenamefont {Freyssingeas},
  \citenamefont {Li\'enard}, \citenamefont {Karpati}, \citenamefont {Parola},
  \citenamefont {Jaurand}, \citenamefont {Dugas}, \citenamefont {Matthews},
  \citenamefont {Gibaud}, \citenamefont {Divoux},\ and\ \citenamefont
  {Manneville}}]{Sudreau:2023}%
  \BibitemOpen
  \bibfield  {author} {\bibinfo {author} {\bibfnamefont {I.}~\bibnamefont
  {Sudreau}}, \bibinfo {author} {\bibfnamefont {M.}~\bibnamefont {Servel}},
  \bibinfo {author} {\bibfnamefont {E.}~\bibnamefont {Freyssingeas}}, \bibinfo
  {author} {\bibfnamefont {F.~m.~c.}\ \bibnamefont {Li\'enard}}, \bibinfo
  {author} {\bibfnamefont {S.}~\bibnamefont {Karpati}}, \bibinfo {author}
  {\bibfnamefont {S.}~\bibnamefont {Parola}}, \bibinfo {author} {\bibfnamefont
  {X.}~\bibnamefont {Jaurand}}, \bibinfo {author} {\bibfnamefont {P.-Y.}\
  \bibnamefont {Dugas}}, \bibinfo {author} {\bibfnamefont {L.}~\bibnamefont
  {Matthews}}, \bibinfo {author} {\bibfnamefont {T.}~\bibnamefont {Gibaud}},
  \bibinfo {author} {\bibfnamefont {T.}~\bibnamefont {Divoux}},\ and\ \bibinfo
  {author} {\bibfnamefont {S.}~\bibnamefont {Manneville}},\ }\bibfield  {title}
  {\bibinfo {title} {Shear-induced stiffening in boehmite gels: A
  rheo-x-ray-scattering study},\ }\href@noop {} {\bibfield  {journal} {\bibinfo
   {journal} {Phys. Rev. Mater.}\ }\textbf {\bibinfo {volume} {7}},\ \bibinfo
  {pages} {115603} (\bibinfo {year} {2023})}\BibitemShut {NoStop}%
\bibitem [{\citenamefont {Dag{\`e}s}\ \emph {et~al.}(2022)\citenamefont
  {Dag{\`e}s}, \citenamefont {Bouthier}, \citenamefont {Matthews},
  \citenamefont {Manneville}, \citenamefont {Divoux}, \citenamefont
  {Poulesquen},\ and\ \citenamefont {Gibaud}}]{Dages:2022}%
  \BibitemOpen
  \bibfield  {author} {\bibinfo {author} {\bibfnamefont {N.}~\bibnamefont
  {Dag{\`e}s}}, \bibinfo {author} {\bibfnamefont {L.~V.}\ \bibnamefont
  {Bouthier}}, \bibinfo {author} {\bibfnamefont {L.}~\bibnamefont {Matthews}},
  \bibinfo {author} {\bibfnamefont {S.}~\bibnamefont {Manneville}}, \bibinfo
  {author} {\bibfnamefont {T.}~\bibnamefont {Divoux}}, \bibinfo {author}
  {\bibfnamefont {A.}~\bibnamefont {Poulesquen}},\ and\ \bibinfo {author}
  {\bibfnamefont {T.}~\bibnamefont {Gibaud}},\ }\bibfield  {title} {\bibinfo
  {title} {Interpenetration of fractal clusters drives elasticity in colloidal
  gels formed upon flow cessation},\ }\href@noop {} {\bibfield  {journal}
  {\bibinfo  {journal} {Soft Matter}\ }\textbf {\bibinfo {volume} {18}},\
  \bibinfo {pages} {6645} (\bibinfo {year} {2022})}\BibitemShut {NoStop}%
\bibitem [{\citenamefont {Sudreau}\ \emph
  {et~al.}(2022{\natexlab{b}})\citenamefont {Sudreau}, \citenamefont {Auxois},
  \citenamefont {Servel}, \citenamefont {L\'ecolier}, \citenamefont
  {Manneville},\ and\ \citenamefont {Divoux}}]{Sudreau:2022b}%
  \BibitemOpen
  \bibfield  {author} {\bibinfo {author} {\bibfnamefont {I.}~\bibnamefont
  {Sudreau}}, \bibinfo {author} {\bibfnamefont {M.}~\bibnamefont {Auxois}},
  \bibinfo {author} {\bibfnamefont {M.}~\bibnamefont {Servel}}, \bibinfo
  {author} {\bibfnamefont {E.}~\bibnamefont {L\'ecolier}}, \bibinfo {author}
  {\bibfnamefont {S.}~\bibnamefont {Manneville}},\ and\ \bibinfo {author}
  {\bibfnamefont {T.}~\bibnamefont {Divoux}},\ }\bibfield  {title} {\bibinfo
  {title} {Residual stresses and shear-induced overaging in boehmite gels},\
  }\href@noop {} {\bibfield  {journal} {\bibinfo  {journal} {Phys. Rev.
  Mater.}\ }\textbf {\bibinfo {volume} {6}},\ \bibinfo {pages} {L042601}
  (\bibinfo {year} {2022}{\natexlab{b}})}\BibitemShut {NoStop}%
\bibitem [{\citenamefont {Trompette}\ and\ \citenamefont
  {Meireles}(2003)}]{Trompette:2003}%
  \BibitemOpen
  \bibfield  {author} {\bibinfo {author} {\bibfnamefont {J.}~\bibnamefont
  {Trompette}}\ and\ \bibinfo {author} {\bibfnamefont {M.}~\bibnamefont
  {Meireles}},\ }\bibfield  {title} {\bibinfo {title} {Ion-specific effect on
  the gelation kinetics of concentrated colloidal silica suspensions},\
  }\href@noop {} {\bibfield  {journal} {\bibinfo  {journal} {J. Colloid
  Interface Sci.}\ }\textbf {\bibinfo {volume} {263}},\ \bibinfo {pages} {522}
  (\bibinfo {year} {2003})}\BibitemShut {NoStop}%
\bibitem [{\citenamefont {Tourbin}\ and\ \citenamefont
  {Frances}(2008)}]{Tourbin:2008}%
  \BibitemOpen
  \bibfield  {author} {\bibinfo {author} {\bibfnamefont {M.}~\bibnamefont
  {Tourbin}}\ and\ \bibinfo {author} {\bibfnamefont {C.}~\bibnamefont
  {Frances}},\ }\bibfield  {title} {\bibinfo {title} {Experimental
  characterization and population balance modelling of the dense silica
  suspensions aggregation process},\ }\href@noop {} {\bibfield  {journal}
  {\bibinfo  {journal} {Chem. Eng. Sci.}\ }\textbf {\bibinfo {volume} {63}},\
  \bibinfo {pages} {5239} (\bibinfo {year} {2008})}\BibitemShut {NoStop}%
\bibitem [{\citenamefont {Cao}\ \emph {et~al.}(2010)\citenamefont {Cao},
  \citenamefont {Cummins},\ and\ \citenamefont {Morris}}]{Cao:2010}%
  \BibitemOpen
  \bibfield  {author} {\bibinfo {author} {\bibfnamefont {X.}~\bibnamefont
  {Cao}}, \bibinfo {author} {\bibfnamefont {H.}~\bibnamefont {Cummins}},\ and\
  \bibinfo {author} {\bibfnamefont {J.}~\bibnamefont {Morris}},\ }\bibfield
  {title} {\bibinfo {title} {Structural and rheological evolution of silica
  nanoparticle gels},\ }\href@noop {} {\bibfield  {journal} {\bibinfo
  {journal} {Soft Matter}\ }\textbf {\bibinfo {volume} {6}},\ \bibinfo {pages}
  {5425} (\bibinfo {year} {2010})}\BibitemShut {NoStop}%
\bibitem [{\citenamefont {van~der Linden}\ \emph {et~al.}(2015)\citenamefont
  {van~der Linden}, \citenamefont {Conch{\'u}ir}, \citenamefont {Spigone},
  \citenamefont {Niranjan}, \citenamefont {Zaccone},\ and\ \citenamefont
  {Cicuta}}]{Vanderlinden:2015}%
  \BibitemOpen
  \bibfield  {author} {\bibinfo {author} {\bibfnamefont {M.}~\bibnamefont
  {van~der Linden}}, \bibinfo {author} {\bibfnamefont {B.~O.}\ \bibnamefont
  {Conch{\'u}ir}}, \bibinfo {author} {\bibfnamefont {E.}~\bibnamefont
  {Spigone}}, \bibinfo {author} {\bibfnamefont {A.}~\bibnamefont {Niranjan}},
  \bibinfo {author} {\bibfnamefont {A.}~\bibnamefont {Zaccone}},\ and\ \bibinfo
  {author} {\bibfnamefont {P.}~\bibnamefont {Cicuta}},\ }\bibfield  {title}
  {\bibinfo {title} {Microscopic origin of the {H}ofmeister effect in gelation
  kinetics of colloidal silica},\ }\href@noop {} {\bibfield  {journal}
  {\bibinfo  {journal} {J. Phys. Chem. Lett.}\ }\textbf {\bibinfo {volume}
  {6}},\ \bibinfo {pages} {2881} (\bibinfo {year} {2015})}\BibitemShut
  {NoStop}%
\bibitem [{\citenamefont {S{\"o}gaard}\ \emph {et~al.}(2018)\citenamefont
  {S{\"o}gaard}, \citenamefont {Funehag},\ and\ \citenamefont
  {Abbas}}]{Sogaard:2018}%
  \BibitemOpen
  \bibfield  {author} {\bibinfo {author} {\bibfnamefont {C.}~\bibnamefont
  {S{\"o}gaard}}, \bibinfo {author} {\bibfnamefont {J.}~\bibnamefont
  {Funehag}},\ and\ \bibinfo {author} {\bibfnamefont {Z.}~\bibnamefont
  {Abbas}},\ }\bibfield  {title} {\bibinfo {title} {Silica sol as grouting
  material: a physio-chemical analysis},\ }\href@noop {} {\bibfield  {journal}
  {\bibinfo  {journal} {Nano Converg.}\ }\textbf {\bibinfo {volume} {5}},\
  \bibinfo {pages} {6} (\bibinfo {year} {2018})}\BibitemShut {NoStop}%
\bibitem [{\citenamefont {Bonacci}\ \emph {et~al.}(2020)\citenamefont
  {Bonacci}, \citenamefont {Chateau}, \citenamefont {Furst}, \citenamefont
  {Fusier}, \citenamefont {Goyon},\ and\ \citenamefont
  {Lema{\^\i}tre}}]{Bonacci:2020}%
  \BibitemOpen
  \bibfield  {author} {\bibinfo {author} {\bibfnamefont {F.}~\bibnamefont
  {Bonacci}}, \bibinfo {author} {\bibfnamefont {X.}~\bibnamefont {Chateau}},
  \bibinfo {author} {\bibfnamefont {E.~M.}\ \bibnamefont {Furst}}, \bibinfo
  {author} {\bibfnamefont {J.}~\bibnamefont {Fusier}}, \bibinfo {author}
  {\bibfnamefont {J.}~\bibnamefont {Goyon}},\ and\ \bibinfo {author}
  {\bibfnamefont {A.}~\bibnamefont {Lema{\^\i}tre}},\ }\bibfield  {title}
  {\bibinfo {title} {Contact and macroscopic ageing in colloidal suspensions},\
  }\href@noop {} {\bibfield  {journal} {\bibinfo  {journal} {Nat. Mater.}\
  }\textbf {\bibinfo {volume} {19}},\ \bibinfo {pages} {775} (\bibinfo {year}
  {2020})}\BibitemShut {NoStop}%
\bibitem [{\citenamefont {Bonacci}\ \emph {et~al.}(2022)\citenamefont
  {Bonacci}, \citenamefont {Chateau}, \citenamefont {Furst}, \citenamefont
  {Goyon},\ and\ \citenamefont {Lema\^{\i}tre}}]{Bonacci:2022}%
  \BibitemOpen
  \bibfield  {author} {\bibinfo {author} {\bibfnamefont {F.}~\bibnamefont
  {Bonacci}}, \bibinfo {author} {\bibfnamefont {X.}~\bibnamefont {Chateau}},
  \bibinfo {author} {\bibfnamefont {E.~M.}\ \bibnamefont {Furst}}, \bibinfo
  {author} {\bibfnamefont {J.}~\bibnamefont {Goyon}},\ and\ \bibinfo {author}
  {\bibfnamefont {A.}~\bibnamefont {Lema\^{\i}tre}},\ }\bibfield  {title}
  {\bibinfo {title} {Yield stress aging in attractive colloidal suspensions},\
  }\href@noop {} {\bibfield  {journal} {\bibinfo  {journal} {Phys. Rev. Lett.}\
  }\textbf {\bibinfo {volume} {128}},\ \bibinfo {pages} {018003} (\bibinfo
  {year} {2022})}\BibitemShut {NoStop}%
\bibitem [{\citenamefont {Bonfanti}\ \emph {et~al.}(2020)\citenamefont
  {Bonfanti}, \citenamefont {Kaplan}, \citenamefont {Charras},\ and\
  \citenamefont {Kabla}}]{Bonfanti:2020}%
  \BibitemOpen
  \bibfield  {author} {\bibinfo {author} {\bibfnamefont {A.}~\bibnamefont
  {Bonfanti}}, \bibinfo {author} {\bibfnamefont {J.~L.}\ \bibnamefont
  {Kaplan}}, \bibinfo {author} {\bibfnamefont {G.}~\bibnamefont {Charras}},\
  and\ \bibinfo {author} {\bibfnamefont {A.}~\bibnamefont {Kabla}},\ }\bibfield
   {title} {\bibinfo {title} {Fractional viscoelastic models for power-law
  materials},\ }\href@noop {} {\bibfield  {journal} {\bibinfo  {journal} {Soft
  Matter}\ }\textbf {\bibinfo {volume} {16}},\ \bibinfo {pages} {6002}
  (\bibinfo {year} {2020})}\BibitemShut {NoStop}%
\bibitem [{\citenamefont {Jaishankar}\ and\ \citenamefont
  {McKinley}(2013)}]{Jaishankar:2013}%
  \BibitemOpen
  \bibfield  {author} {\bibinfo {author} {\bibfnamefont {A.}~\bibnamefont
  {Jaishankar}}\ and\ \bibinfo {author} {\bibfnamefont {G.~H.}\ \bibnamefont
  {McKinley}},\ }\bibfield  {title} {\bibinfo {title} {Power-law rheology in
  the bulk and at the interface: quasi-properties and fractional constitutive
  equations},\ }\href@noop {} {\bibfield  {journal} {\bibinfo  {journal} {Proc.
  R. Soc. A: Math. Phys. Eng. Sci.}\ }\textbf {\bibinfo {volume} {469}},\
  \bibinfo {pages} {20120284} (\bibinfo {year} {2013})}\BibitemShut {NoStop}%
\bibitem [{SM()}]{SM}%
  \BibitemOpen
  \href@noop {} {}\bibinfo {note} {See Supplemental Material at [URL] for
  Movies corresponding to Figures~6, 7, and 15.}\BibitemShut {Stop}%
\bibitem [{\citenamefont {Ewoldt}\ \emph {et~al.}(2008)\citenamefont {Ewoldt},
  \citenamefont {Hosoi},\ and\ \citenamefont {McKinley}}]{Ewoldt:2008}%
  \BibitemOpen
  \bibfield  {author} {\bibinfo {author} {\bibfnamefont {R.~H.}\ \bibnamefont
  {Ewoldt}}, \bibinfo {author} {\bibfnamefont {A.}~\bibnamefont {Hosoi}},\ and\
  \bibinfo {author} {\bibfnamefont {G.~H.}\ \bibnamefont {McKinley}},\
  }\bibfield  {title} {\bibinfo {title} {New measures for characterizing
  nonlinear viscoelasticity in large amplitude oscillatory shear},\ }\href@noop
  {} {\bibfield  {journal} {\bibinfo  {journal} {J. Rheol.}\ }\textbf {\bibinfo
  {volume} {52}},\ \bibinfo {pages} {1427} (\bibinfo {year}
  {2008})}\BibitemShut {NoStop}%
\bibitem [{\citenamefont {Ewoldt}\ \emph {et~al.}(2010)\citenamefont {Ewoldt},
  \citenamefont {Winter}, \citenamefont {Maxey},\ and\ \citenamefont
  {McKinley}}]{Ewoldt:2010}%
  \BibitemOpen
  \bibfield  {author} {\bibinfo {author} {\bibfnamefont {R.~H.}\ \bibnamefont
  {Ewoldt}}, \bibinfo {author} {\bibfnamefont {P.}~\bibnamefont {Winter}},
  \bibinfo {author} {\bibfnamefont {J.}~\bibnamefont {Maxey}},\ and\ \bibinfo
  {author} {\bibfnamefont {G.~H.}\ \bibnamefont {McKinley}},\ }\bibfield
  {title} {\bibinfo {title} {Large amplitude oscillatory shear of pseudoplastic
  and elastoviscoplastic materials},\ }\href@noop {} {\bibfield  {journal}
  {\bibinfo  {journal} {Rheol. Acta}\ }\textbf {\bibinfo {volume} {49}},\
  \bibinfo {pages} {191} (\bibinfo {year} {2010})}\BibitemShut {NoStop}%
\bibitem [{\citenamefont {Hyun}\ \emph {et~al.}(2002)\citenamefont {Hyun},
  \citenamefont {Kim}, \citenamefont {Ahn},\ and\ \citenamefont
  {Lee}}]{Hyun:2002}%
  \BibitemOpen
  \bibfield  {author} {\bibinfo {author} {\bibfnamefont {K.}~\bibnamefont
  {Hyun}}, \bibinfo {author} {\bibfnamefont {S.~H.}\ \bibnamefont {Kim}},
  \bibinfo {author} {\bibfnamefont {K.~H.}\ \bibnamefont {Ahn}},\ and\ \bibinfo
  {author} {\bibfnamefont {S.~J.}\ \bibnamefont {Lee}},\ }\bibfield  {title}
  {\bibinfo {title} {Large amplitude oscillatory shear as a way to classify the
  complex fluids},\ }\href@noop {} {\bibfield  {journal} {\bibinfo  {journal}
  {J. Non-Newton. Fluid Mech.}\ }\textbf {\bibinfo {volume} {107}},\ \bibinfo
  {pages} {51} (\bibinfo {year} {2002})}\BibitemShut {NoStop}%
\bibitem [{\citenamefont {Hyun}\ \emph {et~al.}(2011)\citenamefont {Hyun},
  \citenamefont {Wilhelm}, \citenamefont {Klein}, \citenamefont {Cho},
  \citenamefont {Nam}, \citenamefont {Ahn}, \citenamefont {Lee}, \citenamefont
  {Ewoldt},\ and\ \citenamefont {McKinley}}]{Hyun:2011}%
  \BibitemOpen
  \bibfield  {author} {\bibinfo {author} {\bibfnamefont {K.}~\bibnamefont
  {Hyun}}, \bibinfo {author} {\bibfnamefont {M.}~\bibnamefont {Wilhelm}},
  \bibinfo {author} {\bibfnamefont {C.~O.}\ \bibnamefont {Klein}}, \bibinfo
  {author} {\bibfnamefont {K.~S.}\ \bibnamefont {Cho}}, \bibinfo {author}
  {\bibfnamefont {J.~G.}\ \bibnamefont {Nam}}, \bibinfo {author} {\bibfnamefont
  {K.~H.}\ \bibnamefont {Ahn}}, \bibinfo {author} {\bibfnamefont {S.~J.}\
  \bibnamefont {Lee}}, \bibinfo {author} {\bibfnamefont {R.~H.}\ \bibnamefont
  {Ewoldt}},\ and\ \bibinfo {author} {\bibfnamefont {G.~H.}\ \bibnamefont
  {McKinley}},\ }\bibfield  {title} {\bibinfo {title} {A review of nonlinear
  oscillatory shear tests: Analysis and application of large amplitude
  oscillatory shear ({LAOS})},\ }\href@noop {} {\bibfield  {journal} {\bibinfo
  {journal} {Prog. Polym. Sci.}\ }\textbf {\bibinfo {volume} {36}},\ \bibinfo
  {pages} {1697} (\bibinfo {year} {2011})}\BibitemShut {NoStop}%
\bibitem [{\citenamefont {Divoux}\ \emph {et~al.}(2024)\citenamefont {Divoux},
  \citenamefont {Agoritsas}, \citenamefont {Aime}, \citenamefont {Barentin},
  \citenamefont {Barrat}, \citenamefont {Benzi}, \citenamefont {Berthier},
  \citenamefont {Bi}, \citenamefont {Biroli}, \citenamefont {Bonn},
  \citenamefont {Bourrianne}, \citenamefont {Bouzid}, \citenamefont {Del~Gado},
  \citenamefont {Delanoë-Ayari}, \citenamefont {Farain}, \citenamefont
  {Fielding}, \citenamefont {Fuchs}, \citenamefont {van~der Gucht},
  \citenamefont {Henkes}, \citenamefont {Jalaal}, \citenamefont {Joshi},
  \citenamefont {Lemaître}, \citenamefont {Leheny}, \citenamefont
  {Manneville}, \citenamefont {Martens}, \citenamefont {Poon}, \citenamefont
  {Popović}, \citenamefont {Procaccia}, \citenamefont {Ramos}, \citenamefont
  {Richards}, \citenamefont {Rogers}, \citenamefont {Rossi}, \citenamefont
  {Sbragaglia}, \citenamefont {Tarjus}, \citenamefont {Toschi}, \citenamefont
  {Trappe}, \citenamefont {Vermant}, \citenamefont {Wyart}, \citenamefont
  {Zamponi},\ and\ \citenamefont {Zare}}]{Divoux:2024}%
  \BibitemOpen
  \bibfield  {author} {\bibinfo {author} {\bibfnamefont {T.}~\bibnamefont
  {Divoux}}, \bibinfo {author} {\bibfnamefont {E.}~\bibnamefont {Agoritsas}},
  \bibinfo {author} {\bibfnamefont {S.}~\bibnamefont {Aime}}, \bibinfo {author}
  {\bibfnamefont {C.}~\bibnamefont {Barentin}}, \bibinfo {author}
  {\bibfnamefont {J.-L.}\ \bibnamefont {Barrat}}, \bibinfo {author}
  {\bibfnamefont {R.}~\bibnamefont {Benzi}}, \bibinfo {author} {\bibfnamefont
  {L.}~\bibnamefont {Berthier}}, \bibinfo {author} {\bibfnamefont
  {D.}~\bibnamefont {Bi}}, \bibinfo {author} {\bibfnamefont {G.}~\bibnamefont
  {Biroli}}, \bibinfo {author} {\bibfnamefont {D.}~\bibnamefont {Bonn}},
  \bibinfo {author} {\bibfnamefont {P.}~\bibnamefont {Bourrianne}}, \bibinfo
  {author} {\bibfnamefont {M.}~\bibnamefont {Bouzid}}, \bibinfo {author}
  {\bibfnamefont {E.}~\bibnamefont {Del~Gado}}, \bibinfo {author}
  {\bibfnamefont {H.}~\bibnamefont {Delanoë-Ayari}}, \bibinfo {author}
  {\bibfnamefont {K.}~\bibnamefont {Farain}}, \bibinfo {author} {\bibfnamefont
  {S.}~\bibnamefont {Fielding}}, \bibinfo {author} {\bibfnamefont
  {M.}~\bibnamefont {Fuchs}}, \bibinfo {author} {\bibfnamefont
  {J.}~\bibnamefont {van~der Gucht}}, \bibinfo {author} {\bibfnamefont
  {S.}~\bibnamefont {Henkes}}, \bibinfo {author} {\bibfnamefont
  {M.}~\bibnamefont {Jalaal}}, \bibinfo {author} {\bibfnamefont {Y.~M.}\
  \bibnamefont {Joshi}}, \bibinfo {author} {\bibfnamefont {A.}~\bibnamefont
  {Lemaître}}, \bibinfo {author} {\bibfnamefont {R.~L.}\ \bibnamefont
  {Leheny}}, \bibinfo {author} {\bibfnamefont {S.}~\bibnamefont {Manneville}},
  \bibinfo {author} {\bibfnamefont {K.}~\bibnamefont {Martens}}, \bibinfo
  {author} {\bibfnamefont {W.~C.~K.}\ \bibnamefont {Poon}}, \bibinfo {author}
  {\bibfnamefont {M.}~\bibnamefont {Popović}}, \bibinfo {author}
  {\bibfnamefont {I.}~\bibnamefont {Procaccia}}, \bibinfo {author}
  {\bibfnamefont {L.}~\bibnamefont {Ramos}}, \bibinfo {author} {\bibfnamefont
  {J.~A.}\ \bibnamefont {Richards}}, \bibinfo {author} {\bibfnamefont
  {S.}~\bibnamefont {Rogers}}, \bibinfo {author} {\bibfnamefont
  {S.}~\bibnamefont {Rossi}}, \bibinfo {author} {\bibfnamefont
  {M.}~\bibnamefont {Sbragaglia}}, \bibinfo {author} {\bibfnamefont
  {G.}~\bibnamefont {Tarjus}}, \bibinfo {author} {\bibfnamefont
  {F.}~\bibnamefont {Toschi}}, \bibinfo {author} {\bibfnamefont
  {V.}~\bibnamefont {Trappe}}, \bibinfo {author} {\bibfnamefont
  {J.}~\bibnamefont {Vermant}}, \bibinfo {author} {\bibfnamefont
  {M.}~\bibnamefont {Wyart}}, \bibinfo {author} {\bibfnamefont
  {F.}~\bibnamefont {Zamponi}},\ and\ \bibinfo {author} {\bibfnamefont
  {D.}~\bibnamefont {Zare}},\ }\bibfield  {title} {\bibinfo {title}
  {Ductile-to-brittle transition and yielding in soft amorphous materials:
  perspectives and open questions},\ }\href@noop {} {\bibfield  {journal}
  {\bibinfo  {journal} {Soft Matter}\ }\textbf {\bibinfo {volume} {20}},\
  \bibinfo {pages} {6868} (\bibinfo {year} {2024})}\BibitemShut {NoStop}%
\bibitem [{\citenamefont {Kamani}\ and\ \citenamefont
  {Rogers}(2024)}]{Kamani:2024}%
  \BibitemOpen
  \bibfield  {author} {\bibinfo {author} {\bibfnamefont {K.~M.}\ \bibnamefont
  {Kamani}}\ and\ \bibinfo {author} {\bibfnamefont {S.~A.}\ \bibnamefont
  {Rogers}},\ }\bibfield  {title} {\bibinfo {title} {Brittle and ductile
  yielding in soft materials},\ }\href@noop {} {\bibfield  {journal} {\bibinfo
  {journal} {Proc. Natl. Acad. Sci. U.S.A.}\ }\textbf {\bibinfo {volume}
  {121}},\ \bibinfo {pages} {e2401409121} (\bibinfo {year} {2024})}\BibitemShut
  {NoStop}%
\bibitem [{\citenamefont {Miyazaki}\ \emph {et~al.}(2006)\citenamefont
  {Miyazaki}, \citenamefont {Wyss}, \citenamefont {Weitz},\ and\ \citenamefont
  {Reichman}}]{Miyazaki:2006}%
  \BibitemOpen
  \bibfield  {author} {\bibinfo {author} {\bibfnamefont {K.}~\bibnamefont
  {Miyazaki}}, \bibinfo {author} {\bibfnamefont {H.~M.}\ \bibnamefont {Wyss}},
  \bibinfo {author} {\bibfnamefont {D.~A.}\ \bibnamefont {Weitz}},\ and\
  \bibinfo {author} {\bibfnamefont {D.~R.}\ \bibnamefont {Reichman}},\
  }\bibfield  {title} {\bibinfo {title} {Nonlinear viscoelasticity of
  metastable complex fluids},\ }\href@noop {} {\bibfield  {journal} {\bibinfo
  {journal} {Europhys. Lett.}\ }\textbf {\bibinfo {volume} {75}},\ \bibinfo
  {pages} {915} (\bibinfo {year} {2006})}\BibitemShut {NoStop}%
\bibitem [{\citenamefont {Wyss}\ \emph {et~al.}(2007)\citenamefont {Wyss},
  \citenamefont {Miyazaki}, \citenamefont {Mattsson}, \citenamefont {Hu},
  \citenamefont {Reichman},\ and\ \citenamefont {Weitz}}]{Wyss:2007}%
  \BibitemOpen
  \bibfield  {author} {\bibinfo {author} {\bibfnamefont {H.~M.}\ \bibnamefont
  {Wyss}}, \bibinfo {author} {\bibfnamefont {K.}~\bibnamefont {Miyazaki}},
  \bibinfo {author} {\bibfnamefont {J.}~\bibnamefont {Mattsson}}, \bibinfo
  {author} {\bibfnamefont {Z.}~\bibnamefont {Hu}}, \bibinfo {author}
  {\bibfnamefont {D.~R.}\ \bibnamefont {Reichman}},\ and\ \bibinfo {author}
  {\bibfnamefont {D.~A.}\ \bibnamefont {Weitz}},\ }\bibfield  {title} {\bibinfo
  {title} {Strain-rate frequency superposition: A rheological probe of
  structural relaxation in soft materials},\ }\href@noop {} {\bibfield
  {journal} {\bibinfo  {journal} {Phys. Rev. Lett.}\ }\textbf {\bibinfo
  {volume} {98}},\ \bibinfo {pages} {238303} (\bibinfo {year}
  {2007})}\BibitemShut {NoStop}%
\bibitem [{\citenamefont {Homer}\ \emph {et~al.}(2016)\citenamefont {Homer},
  \citenamefont {Lundin},\ and\ \citenamefont {Dunstan}}]{Homer:2016}%
  \BibitemOpen
  \bibfield  {author} {\bibinfo {author} {\bibfnamefont {S.}~\bibnamefont
  {Homer}}, \bibinfo {author} {\bibfnamefont {L.}~\bibnamefont {Lundin}},\ and\
  \bibinfo {author} {\bibfnamefont {D.~E.}\ \bibnamefont {Dunstan}},\
  }\bibfield  {title} {\bibinfo {title} {Modifying the microstructure and
  mechanical properties of whey protein isolate gels using large deformation
  oscillatory strain},\ }\href@noop {} {\bibfield  {journal} {\bibinfo
  {journal} {Food Hydrocoll.}\ }\textbf {\bibinfo {volume} {61}},\ \bibinfo
  {pages} {672} (\bibinfo {year} {2016})}\BibitemShut {NoStop}%
\bibitem [{\citenamefont {Altmann}\ \emph {et~al.}(2004)\citenamefont
  {Altmann}, \citenamefont {Cooper-White}, \citenamefont {Dunstan},\ and\
  \citenamefont {Stokes}}]{Altmann:2004}%
  \BibitemOpen
  \bibfield  {author} {\bibinfo {author} {\bibfnamefont {N.}~\bibnamefont
  {Altmann}}, \bibinfo {author} {\bibfnamefont {J.}~\bibnamefont
  {Cooper-White}}, \bibinfo {author} {\bibfnamefont {D.}~\bibnamefont
  {Dunstan}},\ and\ \bibinfo {author} {\bibfnamefont {J.}~\bibnamefont
  {Stokes}},\ }\bibfield  {title} {\bibinfo {title} {Strong through to weak
  ‘sheared’ gels},\ }\href@noop {} {\bibfield  {journal} {\bibinfo
  {journal} {J. Non-Newton. Fluid Mech.}\ }\textbf {\bibinfo {volume} {124}},\
  \bibinfo {pages} {129} (\bibinfo {year} {2004})}\BibitemShut {NoStop}%
\bibitem [{\citenamefont {Stokes}\ and\ \citenamefont
  {Frith}(2008)}]{Stokes:2008}%
  \BibitemOpen
  \bibfield  {author} {\bibinfo {author} {\bibfnamefont {J.~R.}\ \bibnamefont
  {Stokes}}\ and\ \bibinfo {author} {\bibfnamefont {W.~J.}\ \bibnamefont
  {Frith}},\ }\bibfield  {title} {\bibinfo {title} {Rheology of gelling and
  yielding soft matter systems},\ }\href@noop {} {\bibfield  {journal}
  {\bibinfo  {journal} {Soft Matter}\ }\textbf {\bibinfo {volume} {4}},\
  \bibinfo {pages} {1133} (\bibinfo {year} {2008})}\BibitemShut {NoStop}%
\bibitem [{\citenamefont {Nelson}\ \emph {et~al.}(2022)\citenamefont {Nelson},
  \citenamefont {Wang}, \citenamefont {Wang}, \citenamefont {Margotta},
  \citenamefont {Sammler}, \citenamefont {Izmitli}, \citenamefont {Katz},
  \citenamefont {Curtis-Fisk}, \citenamefont {Li},\ and\ \citenamefont
  {Ewoldt}}]{Nelson:2022}%
  \BibitemOpen
  \bibfield  {author} {\bibinfo {author} {\bibfnamefont {A.~Z.}\ \bibnamefont
  {Nelson}}, \bibinfo {author} {\bibfnamefont {Y.}~\bibnamefont {Wang}},
  \bibinfo {author} {\bibfnamefont {Y.}~\bibnamefont {Wang}}, \bibinfo {author}
  {\bibfnamefont {A.~S.}\ \bibnamefont {Margotta}}, \bibinfo {author}
  {\bibfnamefont {R.~L.}\ \bibnamefont {Sammler}}, \bibinfo {author}
  {\bibfnamefont {A.}~\bibnamefont {Izmitli}}, \bibinfo {author} {\bibfnamefont
  {J.~S.}\ \bibnamefont {Katz}}, \bibinfo {author} {\bibfnamefont
  {J.}~\bibnamefont {Curtis-Fisk}}, \bibinfo {author} {\bibfnamefont
  {Y.}~\bibnamefont {Li}},\ and\ \bibinfo {author} {\bibfnamefont {R.~H.}\
  \bibnamefont {Ewoldt}},\ }\bibfield  {title} {\bibinfo {title} {Gelation
  under stress: Impact of shear flow on the formation and mechanical properties
  of methylcellulose hydrogels},\ }\href@noop {} {\bibfield  {journal}
  {\bibinfo  {journal} {Soft Matter}\ }\textbf {\bibinfo {volume} {18}},\
  \bibinfo {pages} {1554} (\bibinfo {year} {2022})}\BibitemShut {NoStop}%
\bibitem [{\citenamefont {Creton}\ and\ \citenamefont
  {Ciccotti}(2016)}]{Creton:2016}%
  \BibitemOpen
  \bibfield  {author} {\bibinfo {author} {\bibfnamefont {C.}~\bibnamefont
  {Creton}}\ and\ \bibinfo {author} {\bibfnamefont {M.}~\bibnamefont
  {Ciccotti}},\ }\bibfield  {title} {\bibinfo {title} {Fracture and adhesion of
  soft materials: a review},\ }\href@noop {} {\bibfield  {journal} {\bibinfo
  {journal} {Rep. Prog. Phys.}\ }\textbf {\bibinfo {volume} {79}},\ \bibinfo
  {pages} {046601} (\bibinfo {year} {2016})}\BibitemShut {NoStop}%
\bibitem [{\citenamefont {van Doorn}\ \emph {et~al.}(2018)\citenamefont {van
  Doorn}, \citenamefont {Verweij}, \citenamefont {Sprakel},\ and\ \citenamefont
  {van~der Gucht}}]{vanDoorm2018}%
  \BibitemOpen
  \bibfield  {author} {\bibinfo {author} {\bibfnamefont {J.~M.}\ \bibnamefont
  {van Doorn}}, \bibinfo {author} {\bibfnamefont {J.~E.}\ \bibnamefont
  {Verweij}}, \bibinfo {author} {\bibfnamefont {J.}~\bibnamefont {Sprakel}},\
  and\ \bibinfo {author} {\bibfnamefont {J.}~\bibnamefont {van~der Gucht}},\
  }\bibfield  {title} {\bibinfo {title} {Strand plasticity governs fatigue in
  colloidal gels},\ }\href@noop {} {\bibfield  {journal} {\bibinfo  {journal}
  {Phys. Rev. Lett.}\ }\textbf {\bibinfo {volume} {120}},\ \bibinfo {pages}
  {208005} (\bibinfo {year} {2018})}\BibitemShut {NoStop}%
\bibitem [{\citenamefont {Verweij}\ \emph {et~al.}(2019)\citenamefont
  {Verweij}, \citenamefont {Leermakers}, \citenamefont {Sprakel},\ and\
  \citenamefont {Van Der~Gucht}}]{Verweij2019}%
  \BibitemOpen
  \bibfield  {author} {\bibinfo {author} {\bibfnamefont {J.~E.}\ \bibnamefont
  {Verweij}}, \bibinfo {author} {\bibfnamefont {F.~A.}\ \bibnamefont
  {Leermakers}}, \bibinfo {author} {\bibfnamefont {J.}~\bibnamefont
  {Sprakel}},\ and\ \bibinfo {author} {\bibfnamefont {J.}~\bibnamefont {Van
  Der~Gucht}},\ }\bibfield  {title} {\bibinfo {title} {Plasticity in colloidal
  gel strands},\ }\href@noop {} {\bibfield  {journal} {\bibinfo  {journal}
  {Soft Matter}\ }\textbf {\bibinfo {volume} {15}},\ \bibinfo {pages} {6447}
  (\bibinfo {year} {2019})}\BibitemShut {NoStop}%
\bibitem [{\citenamefont {Donley}\ \emph {et~al.}(2022)\citenamefont {Donley},
  \citenamefont {Bantawa},\ and\ \citenamefont {Del~Gado}}]{Donley:2022}%
  \BibitemOpen
  \bibfield  {author} {\bibinfo {author} {\bibfnamefont {G.~J.}\ \bibnamefont
  {Donley}}, \bibinfo {author} {\bibfnamefont {M.}~\bibnamefont {Bantawa}},\
  and\ \bibinfo {author} {\bibfnamefont {E.}~\bibnamefont {Del~Gado}},\
  }\bibfield  {title} {\bibinfo {title} {Time-resolved microstructural changes
  in large amplitude oscillatory shear of model single and double component
  soft gels},\ }\href@noop {} {\bibfield  {journal} {\bibinfo  {journal} {J.
  Rheol.}\ }\textbf {\bibinfo {volume} {66}},\ \bibinfo {pages} {1287}
  (\bibinfo {year} {2022})}\BibitemShut {NoStop}%
\bibitem [{\citenamefont {Chaudhury}\ \emph {et~al.}(2015)\citenamefont
  {Chaudhury}, \citenamefont {Chakrabarti},\ and\ \citenamefont
  {Ghatak}}]{Chaudhury:2015}%
  \BibitemOpen
  \bibfield  {author} {\bibinfo {author} {\bibfnamefont {M.~K.}\ \bibnamefont
  {Chaudhury}}, \bibinfo {author} {\bibfnamefont {A.}~\bibnamefont
  {Chakrabarti}},\ and\ \bibinfo {author} {\bibfnamefont {A.}~\bibnamefont
  {Ghatak}},\ }\bibfield  {title} {\bibinfo {title} {Adhesion-induced
  instabilities and pattern formation in thin films of elastomers and gels},\
  }\href@noop {} {\bibfield  {journal} {\bibinfo  {journal} {Eur. Phys. J. E}\
  }\textbf {\bibinfo {volume} {38}},\ \bibinfo {pages} {82} (\bibinfo {year}
  {2015})}\BibitemShut {NoStop}%
\bibitem [{\citenamefont {Fardin}\ \emph {et~al.}(2025)\citenamefont {Fardin},
  \citenamefont {Divoux}, \citenamefont {Lee},\ and\ \citenamefont
  {Bischofberger}}]{Fardin:2025}%
  \BibitemOpen
  \bibfield  {author} {\bibinfo {author} {\bibfnamefont {M.~A.}\ \bibnamefont
  {Fardin}}, \bibinfo {author} {\bibfnamefont {T.}~\bibnamefont {Divoux}},
  \bibinfo {author} {\bibfnamefont {S.}~\bibnamefont {Lee}},\ and\ \bibinfo
  {author} {\bibfnamefont {I.}~\bibnamefont {Bischofberger}},\ }\bibfield
  {title} {\bibinfo {title} {Pattern formation and instabilities in particulate
  suspensions},\ }\href@noop {} {\bibfield  {journal} {\bibinfo  {journal}
  {Annu. Rev. Fluid Mech.}\ }\textbf {\bibinfo {volume} {58}} (\bibinfo {year}
  {2025})}\BibitemShut {NoStop}%
\bibitem [{DAS()}]{DAS}%
  \BibitemOpen
  \href@noop {} {}\bibinfo {note}
  {Https://zenodo.org/records/17755050}\BibitemShut {NoStop}%
\bibitem [{\citenamefont {Narayanan}\ \emph {et~al.}(2022)\citenamefont
  {Narayanan}, \citenamefont {Sztucki}, \citenamefont {Zinn}, \citenamefont
  {Kieffer}, \citenamefont {Homs-Puron}, \citenamefont {Gorini}, \citenamefont
  {Van~Vaerenbergh},\ and\ \citenamefont {Boesecke}}]{Narayanan:2022}%
  \BibitemOpen
  \bibfield  {author} {\bibinfo {author} {\bibfnamefont {T.}~\bibnamefont
  {Narayanan}}, \bibinfo {author} {\bibfnamefont {M.}~\bibnamefont {Sztucki}},
  \bibinfo {author} {\bibfnamefont {T.}~\bibnamefont {Zinn}}, \bibinfo {author}
  {\bibfnamefont {J.}~\bibnamefont {Kieffer}}, \bibinfo {author} {\bibfnamefont
  {A.}~\bibnamefont {Homs-Puron}}, \bibinfo {author} {\bibfnamefont
  {J.}~\bibnamefont {Gorini}}, \bibinfo {author} {\bibfnamefont
  {P.}~\bibnamefont {Van~Vaerenbergh}},\ and\ \bibinfo {author} {\bibfnamefont
  {P.}~\bibnamefont {Boesecke}},\ }\bibfield  {title} {\bibinfo {title}
  {Performance of the time-resolved ultra-small-angle {X}-ray scattering
  beamline with the extremely brilliant source},\ }\href@noop {} {\bibfield
  {journal} {\bibinfo  {journal} {J. Appl. Crystallogr.}\ }\textbf {\bibinfo
  {volume} {55}},\ \bibinfo {pages} {98} (\bibinfo {year} {2022})}\BibitemShut
  {NoStop}%
\end{thebibliography}

%

\end{document}